\documentclass[11pt, a4paper]{article}
\usepackage{jheppub}

\usepackage[utf8]{inputenc}

\usepackage{bm}
\usepackage{comment} 
\usepackage{mathtools}

\begin{document}

\title{The First Law and Weak Cosmic Censorship for de~Sitter Black Holes}

\date{\today}

\author[a]{Daisuke Yoshida}
\affiliation[a]{Department of Mathematics, Nagoya University, Nagoya 464-8602, Japan}
\emailAdd{dyoshida@math.nagoya-u.ac.jp}

\author[b]{and Kaho Yoshimura}
\affiliation[b]{Graduate School of Arts and Sciences, University of Tokyo, Komaba, Meguro-ku, Tokyo 153-8902, Japan}
\emailAdd{yoshimura-kaho848@g.ecc.u-tokyo.ac.jp}

\preprint{UT-Komaba/24-9}

\abstract{
We apply Iyer--Wald's covariant phase space formalism to asymptotically de Sitter spacetimes and establish the thermodynamic first law, expressed in terms of the Abbott--Deser mass. Similar to Iyer--Wald's first law for asymptotically flat black holes, our first law applies to general asymptotically de Sitter perturbations around a Reissner--Nordstr\"{o}m--de Sitter black hole, without imposing any symmetry for the perturbations. We explicitly derive the first law up to the second order in perturbation. Additionally, we apply this first law to a thought experiment involving the overcharging of a Reissner--Nordstr\"{o}m--de Sitter black hole by injecting energy and charge sources. We find that asymptotically de Sitter black holes cannot be overcharged, provided that the null energy condition holds, thereby reinforcing the weak cosmic censorship conjecture. 
}

\maketitle
\section{Introduction}
The analogy between the laws of black hole mechanics and those of thermodynamics, first clarified by Bardeen, Carter, and Hawking \cite{Bardeen:1973gs}, is important for understanding the nature of black holes.
The covariant phase space formalism, established by Iyer and Wald \cite{Wald:1993nt,Iyer:1994ys,Iyer:1995kg} is a useful formulation of the first law of black hole thermodynamics for arbitrary perturbations around a stationary black hole in any diffeomorphism invariant systems, provided that the spacetime is asymptotically flat. In particular, applying their formalism to the Einstein--Maxwell system, one can obtain the first law relation
\begin{align}
\label{eq:intro_1stlaw}
    \delta M - \Omega_{\mathrm{H}} \delta J - \frac{\kappa}{8\pi G}\delta A_{B} -\Phi_{\mathrm{H}} \delta Q_{B} = 0,
\end{align}
for any perturbations around the Kerr--Newman class of black hole spacetimes, provided that the perturbations satisfy the electro-vacuum Einstein--Maxwell equations of motion.
Here $\delta M$ and $\delta J$ are the perturbation of the Arnowitt, Deser, and Misner (ADM) mass \cite{Arnowitt:1960es} and the Komar angular momentum \cite{Komar:1958wp}, evaluated at the spatial infinity. The subscript $B$ indicates that the quantities are evaluated on the bifurcation surface $B$ of the background black hole spacetime. Specifically, $\delta Q_{B}$ represents the perturbation of the charge contained within $B$, and $\delta A_{B}$ is the perturbation of the area of $B$. Additionally, $\kappa$, $\Omega_{\mathrm{H}}$, and $\Phi_{\mathrm{H}}$ denote the surface gravity, angular velocity, and Coulomb potential respectively, associated with the background spacetime, evaluated at the bifurcation surface $B$. Finally, $G$ represents the Newton's gravitational constant. Note that, throughout the paper, we employ the unit where the speed of light is set to unity.  
 
Since the covariant phase space formalism is established in a non-perturbative manner, it can be applied to any order of perturbations. In fact, the first law for the second order perturbations is studied in Ref.~\cite{Hollands:2012sf} to explore the thermodynamic stability of black hole and black brane solutions to the vacuum Einstein equations. Additionally, since this formalism is established in an off-shell manner, it can also be applied to situations where the perturbations are not solutions to the equations of motion of the system -- that is, situations where external sources to the equations of motion are present.
In this paper, we focus on an interesting application of this higher-order, off-shell version of the first law -- developed by Sorce and Wald \cite{Sorce:2017dst} for asymptotically flat spacetimes -- to the weak cosmic censorship conjecture~\cite{Penrose:1969pc}.
 
The weak cosmic censorship conjecture~\cite{Penrose:1969pc} states that spacetime singularities resulting from gravitational collapse must be hidden within event horizons. While this conjecture has been tested through numerous examples, the conditions under which it holds remain an active area of study. 
One thought experiment along these lines is to investigate whether a black hole can be overcharged or overspun by injecting energy, angular momentum, and charge from the outside, as first proposed in~Ref.~\cite{Wald:1974hkz}. 
A typical analysis considers a test particle falling into a black hole. The goal is to compare the condition for the particle to reach the event horizon with the condition for the total energy of the system to fall below that of an extremal black hole, as determined by the total angular momentum and the charge of the system.
This approach has been applied to various systems, and the common results can be summarized as follows:
The first result is that the extremal black holes cannot be overcharged or overspun by injecting a test particle~\cite{Wald:1974hkz, Cohen:1979zzb, Needham:1980fb, hiscock1981magnetic, Gwak:2015fsa, Duztas:2016xfg, Ghosh:2019dzq, Yang:2020iat, Izumi:2024rge}. The second result is that slightly sub-extremal black holes seem to be able to get overcharged or overspun~\cite{Hubeny:1998ga, deFelice:2001wj, Hod:2002pm, Jacobson:2009kt, Chirco:2010rq, Saa:2011wq, Gao:2012ca, Duztas:2016xfg, Yang:2020iat} if one uses the test particle approximation, that is, by omitting the backreaction from the particle.
 Note that the weak cosmic censorship is tested by injecting fields in Refs.~\cite{Semiz:2005gs, Duztas:2013wua, Duztas:2013gza, Duztas:2014sga, Semiz:2015pna, Duztas:2016xfg, Duztas:2017lxk, Gwak:2018akg, Duztas:2018adf, Duztas:2019ick, Yang:2020iat}.
To conclude whether the second result implies a violation of the weak cosmic censorship conjecture, one needs to treat the backreaction from the injected energy, angular momentum, and charge more seriously. A possible way to incorporate the backreaction is to assume a high symmetry of the system and solve the dynamics non-perturbatively, such as by considering the injection of a spherically symmetric shell \cite{farrugia1979third, hiscock1981magnetic, proszynski1983thin, cruz1967gravitational, Boulware:1973tlq,Giese:2022ljq}.  
The Sorce--Wald formalism~\cite{Sorce:2017dst}, as described above, offers another way to treat such backreaction. In contrast to the analysis of the shell, one does not need to impose any symmetry on the full system, although the analysis needs to be perturbative.
Considering the backreaction from the first-order perturbations, Sorce and Wald show that a slightly sub-extremal Kerr--Newman black hole cannot be overcharged or overspun by injecting any energy, angular momentum, and/or charge, as long as it satisfies the null energy conditions. Thus, the apparent violation of the weak cosmic censorship, shown within the test particle approximation, is prevented by the contribution of the backreaction.
The Sorce--Wald formalism, then, has been applied for various black holes \cite{An:2017phb,Ge:2017vun,Duztas:2018fbc,Chen:2019nhv,Shaymatov:2019del,Shaymatov:2019pmn,Jiang:2019ige,Jiang:2019vww,Wang:2019bml,He:2019mqy,Jiang:2019soz,Jiang:2020btc,Jiang:2020mws,Wang:2020vpn,Shaymatov:2020byu,Jiang:2020alh,Qu:2020nac,Jiang:2020xow,Zhang:2020txy,Li:2020smq,Ding:2020zgg,Duztas:2021kni,Qu:2021hxh,Sang:2021xqj,Huang:2022avq,Sang:2022lfk,Molla:2022dqr,Ahmed:2022dpu,Wang:2022umx,Jiang:2022zod,Yang:2023hll,Duztas:2024vky}, including both asymptotically flat black holes and asymptotically anti de Sitter black holes.
See also Refs.~\cite{Lin:2022ndf, Lin:2024deg, Wu:2024ucf} for another approach to the cosmic censorship conjecture based on the second law.

The weak cosmic censorship conjecture plays a crucial role even also in the 
study of the spectrum of charged particles through thought experiments on black hole evaporation.
For instance, the existence of a charged particle with a mass smaller than the upper bound determined by the charge is
suggested to ensure that extremal black holes can evaporate
without exposing naked singularities, which is known as the weak gravity conjecture \cite{Arkani-Hamed:2006emk}. In this way, 
when studying the particle spectrum based on the properties of gravity,
the weak cosmic censorship conjecture is often regarded as
a fundamental principle.
Thus, 
it is important to clarify under what circumstances the cosmic censorship conjecture holds, even in this context.

In contrast to asymptotically flat or anti de Sitter black holes, the thermodynamic first law for asymptotically de Sitter black holes is not well understood.
An application of Sorce--Wald formalism to an asymptotically de Sitter black hole surrounded by dark matter is discussed by Ref.~\cite{Shaymatov:2020wtj}, formally applying the formula for the asymptotically flat case. 
However validity of the direct application is non-trivial because the spacetime structure of de Sitter spacetimes is different from that of flat spacetimes. 
The main difference is the absence of spatial infinity, where the total mass of the system can be defined, in de Sitter spacetime. We aim to address this issue, drawing inspiration from the studies~\cite{Abbott:1981ff, Shiromizu:1993fi, Nakao:1994fj}, which attempt to define the mass in asymptotically de Sitter spacetimes based on the static Killing vector of the de Sitter space. The mass defined in this manner is referred to as the Abbott--Deser mass. In the original work by Abbott and Deser \cite{Abbott:1981ff}, the mass is evaluated on a sufficiently large co-dimension two surface that encloses the black hole but remains inside the cosmological horizon. Subsequently, Nakao, Shiromizu, and Maeda \cite{Shiromizu:1993fi, Nakao:1994fj} extended the definition of the Abbott--Deser mass to an infinitely large sphere beyond the cosmological horizon, corresponding to the apparent ``spatial infinity'' in the flat chart of de Sitter space. Although the ``static'' Killing vector becomes spacelike there, it has been shown that the Abbott--Deser mass reduces the standard mass parameter when evaluated for the Schwarzschild -- de Sitter spacetime.
Note that a definition of the mass at the future null infinity in de Sitter spacetime is also discussed in Ref.~\cite{Anninos:2010zf}.

In this paper, we apply the covariant phase space formalism to de Sitter black holes. By incorporating the concept of ``spatial infinity'' for de Sitter black holes as described above, we are able to formulate the thermodynamic first law, expressed in terms of the Abbott--Deser mass. It is worth mentioning that similar analysis within the static coordinates is studied in Refs.~\cite{Urano:2009xn, Banihashemi:2022htw}. Additionally, as an application of our first law, we extend the Sorce--Wald formalism to asymptotically de Sitter black holes and confirm that Reissner--Nordstr\"{o}m--de Sitter black holes cannot be overcharged by injecting any energy or charge sources, provided that the null energy condition is satisfied. Thus, our results support the weak cosmic censorship conjecture. 

This paper is organized as follows. In the next section, we review the covariant phase space formalism with an application to the Einstein--Maxwell system. In Sec.~\ref{sec:BHindS}, we explain the definition of ``asymptotic de Sitter spacetime'' and ``spatial infinity''. In Sec.~\ref{Sec:firstlaw}, we apply the covariant phase space formalism to asymptotic de Sitter spacetimes and establish the first law for both first and second order perturbations. In Sec.~\ref{sec:WCC}, we apply our first law to the thought experiment on the weak cosmic censorship conjecture following the Sorce--Wald formalism and confirm that the Reissner--Nordstr\"{o}m--de Sitter black holes cannot be overcharged. The final section is dedicated to the summary and discussion. 
In Appendix \ref{App:GNC}, we present detailed calculations associated with the Gaussian null coordinates \cite{Hollands:2006rj,Hollands:2012sf}. In Appendix \ref{App:2nd}, we discuss the relation between our result and an approach based on the thermodynamic second law~\cite{Lin:2022ndf, Lin:2024deg, Wu:2024ucf}. We find that our result can be reproduced by requiring the second law for the black hole entropy. This contrasts with the findings in Ref.~\cite{Lin:2024deg}, which suggest that the weak cosmic censorship conjecture could be violated when the second law is applied to the total entropy of both the event and cosmological horizons. 

Throughout the paper, we use the bold style notation for the abstract tensors following the textbook by Misner, Thorn, and Wheeler \cite{Misner:1973prb}. 
Additionally, we use short hand notion for the symmetric tensor product as $\boldsymbol{u} \boldsymbol{v} \coloneqq \frac{1}{2} \left( \boldsymbol{u} \otimes \boldsymbol{v} + \boldsymbol{v} \otimes \boldsymbol{u} \right)$. 

\section{Review on the Covariant Phase Space Formalism}
In this section, we review the covariant phase space formalism \cite{Wald:1993nt, Iyer:1994ys, Iyer:1995kg,Sorce:2017dst,Hollands:2012sf}, with a specific application to Einstein--Maxwell system with a cosmological constant. 

The action for a diffeomorphism invariant system in 4-dimensional spacetime can be expressed as the integral of the Lagrangian 4-form $\boldsymbol{\mathcal{L}}$, which is constructed covariantly from the dynamical fields, which we express as $\phi$.
For Einstein--Maxwell system with a cosmological constant $\Lambda$, the Lagrangian 4-form $\boldsymbol{\mathcal{L}}$ is given by 
\begin{align}
 \boldsymbol{\mathcal{L}}(\phi) = \left[
\frac{1}{16 \pi G} ( R - 2 \Lambda )  - \frac{1}{4} F_{\mu\nu} F^{\mu\nu}
\right] \boldsymbol{\epsilon},\label{EinsteinMaxwellLagrangian}
\end{align}
where $R$ is the Ricci scalar associated with the metric tensor $\boldsymbol{g}$, $\boldsymbol{F} = \mathbf{d} \boldsymbol{A}$ is the field strength of the Maxwell field  $\boldsymbol{A}$, and $\boldsymbol{\epsilon}$ is the volume 4-form which can be expressed as 
\begin{align}
\boldsymbol{\epsilon} =  \sqrt{- g} ~ \mathbf{d}x^{0} \wedge \mathbf{d}x^{1}\wedge \mathbf{d}x^{2} \wedge \mathbf{d}x^{3},
\end{align}
in a given set of coordinate basis $\{ \mathbf{d} x^{\mu} \}$. In this paper, Greek indices represent the components in an arbitrarily chosen set of coordinate basis, unless otherwise specified. 
Here $\phi$ represents all the dynamical variables, and in our case, it represents $ \phi  = \{ g_{\mu\nu}, A_{\mu} \} $.

For arbitrary one parameter family of fields configuration $\phi(\lambda)$, the variation of the Lagrangian can be calculated as 
\begin{align}
 \frac{d \boldsymbol{\mathcal{L}}(\phi(\lambda))}{d \lambda}
= \boldsymbol{E}(\phi) \cdot \frac{d \phi}{d \lambda}
+ \mathbf{d} \boldsymbol{\theta}\left( \phi ; \frac{d \phi}{d \lambda}\right),\label{variationofL}
\end{align}
where the 3-form $\boldsymbol{\theta}$ corresponds to the boundary term and the 4-form $\boldsymbol{E}$ is related to the equations of motion derived from the action given by $\boldsymbol{E}(\phi)=0$. The specific forms of $\boldsymbol{E}$ for the given Lagrangian \eqref{EinsteinMaxwellLagrangian} are given as follows;
\begin{align}
 \boldsymbol{E}(\phi) \cdot \frac{d \phi}{d \lambda}
&= - \boldsymbol{\epsilon} \left[ \frac{1}{2} T^{\mu\nu} \frac{d g_{\mu\nu}}{d \lambda} + j^{\mu} \frac{d A_{\mu}}{d \lambda}  \right], 
\end{align}
with defining source terms $T^{\mu\nu}$ and $j^{\mu}$ by
\begin{align}
8 \pi G T^{\mu\nu} &\coloneqq R^{\mu\nu} - \frac{1}{2} g^{\mu\nu} R + \Lambda g^{\mu\nu} - 8 \pi G \left(T^{\mathrm{EM}}\right)^{\mu\nu}, \\
j^{\mu} &\coloneqq \nabla_{\nu} F^{\mu\nu},
\end{align}
and the energy momentum tensor for electromagnetic fields $\left(T^{\mathrm{EM}}\right)^{\mu\nu}$ by
\begin{align}
    \left( T^{\mathrm{EM}}\right)^{\mu\nu}&=F^{\mu\rho} F^{\nu}{}_{\rho}-\frac{1}{4}F_{\rho\sigma}F^{\rho\sigma}g^{\mu\nu}.
\end{align}
The equations of motion correspond to $T_{\mu\nu} = 0$ and $j^{\mu} = 0$.
The boundary term 3-form $\boldsymbol{\theta}$ can be obtained as 
\begin{align}
 \boldsymbol{\theta} \left(\phi, \frac{d\phi}{d\lambda}\right)\coloneqq \boldsymbol{\theta}^{\mathrm{GR}} + \boldsymbol{\theta}^{\mathrm{EM}},
\end{align}
where the components of each terms can be described as 
\begin{align}
 \theta^{\mathrm{GR}}_{\mu\nu\rho}\left(\phi; \frac{d \phi}{d\lambda}\right) &\coloneqq 
\frac{1}{16 \pi G} \epsilon_{\sigma\mu\nu\rho } g^{\sigma \alpha} g^{\beta \gamma}
\left(
\nabla_{\beta} \frac{d g_{\alpha\gamma}}{d \lambda} - \nabla_{\alpha} \frac{d g_{\beta\gamma}}{d \lambda}
\right), \\
 \theta^{\mathrm{EM}}_{\mu\nu\rho}\left(\phi; \frac{d \phi}{d\lambda}\right) &\coloneqq - \epsilon_{\sigma\mu\nu\rho} F^{\sigma\alpha} \frac{d A_{\alpha}}{d \lambda}.
\end{align}
We note that the components of the 3-form are defined as 
\begin{align}
    \boldsymbol{\theta}^{\mathrm{GR}} = \frac{1}{3!} \theta^{\mathrm{GR}}_{\mu\nu\rho}~ \mathbf{d}x^\mu \wedge \mathbf{d}x^\nu \wedge \mathbf{d}x^\rho,
\end{align}
and the same applies to the electromagnetic one. 
Throughout this paper we use this notation methods for differential forms and their components.

Since this system is diffeomorphism invariant, one can confirm that the Lie derivative of the Lagrangian 4-form $\boldsymbol{\mathcal{L}}$ along arbitrary vector field $\boldsymbol{\xi}$ becomes a total derivative,
\begin{align}
 \mathsterling_{\boldsymbol{\xi}} \boldsymbol{\mathcal{L}} = \mathbf{d} i_{\boldsymbol{\xi}} \boldsymbol{\mathcal{L}} + i_{\boldsymbol{\xi}} \mathbf{d} \boldsymbol{\mathcal{L}}  = \mathbf{d} i_{\boldsymbol{\xi}} \boldsymbol{\mathcal{L}}.
\end{align}
Here, $i_{\boldsymbol{\xi}}$ denotes an inner product where the first index of the differential form $\mathcal{L}$ is contracted with $\boldsymbol{\xi}$.
Comparing this expression with the diffeomorphism variation of the Lagrangian  i.e. Eq.~\eqref{variationofL} for Lie derivative, that is,
\begin{align}
 \mathsterling_{\boldsymbol{\xi}} \boldsymbol{\mathcal{L}} = 
\boldsymbol{E}(\phi) \cdot \mathsterling_{\boldsymbol{\xi}} \phi
+ \mathbf{d} \boldsymbol{\theta}\left( \phi ; \mathsterling_{\boldsymbol{\xi}} \phi\right),
\end{align}
we can define the Noether current 3-form $\boldsymbol{\mathcal{J}}_{\boldsymbol{\xi}}$ associated with a vector field $\boldsymbol{\xi}$ by
\begin{align}
 \boldsymbol{\mathcal{J}}_{\boldsymbol{\xi}} \coloneqq \boldsymbol{\theta}\left(\phi; \mathsterling_{\boldsymbol{\xi}} \phi \right) - i_{\boldsymbol{\xi}} \boldsymbol{\mathcal{L}}(\phi) \label{defJ},
\end{align}
and it is conserved on-shell,
\begin{align}
 \mathbf{d} \boldsymbol{\mathcal{J}}_{\boldsymbol{\xi}}  = - \boldsymbol{E} \cdot \mathsterling_{\boldsymbol{\xi}} \phi \overset{\boldsymbol{E} = 0}{=} 0.
\end{align}

Iyer--Wald's identity can be derived by expressing the Noether current 3-form $\boldsymbol{\mathcal{J}}_{\boldsymbol{\xi}}$ in another way. 
According to the general discussion in Ref.~\cite{Wald:1993nt,Iyer:1994ys,Iyer:1995kg}, or simply calculating by using the specific Lagrangian \eqref{EinsteinMaxwellLagrangian}, the Noether current 3-form $\boldsymbol{\mathcal{J}}_{\boldsymbol{\xi}}$ can be expressed as
\begin{align}
\boldsymbol{\mathcal{J}}_{\boldsymbol{\xi}} = \mathbf{d}\boldsymbol{Q}_{\boldsymbol{\xi}} + \boldsymbol{C}_{\boldsymbol{\xi}},\label{J_expressed_by_Q}
\end{align}
where the Noether charge 2-form $\boldsymbol{Q}_{\boldsymbol{\xi}}$ and the source term 3-form $\boldsymbol{C}_{\boldsymbol{\xi}}$ are covariantly constructed from the dynamical variables. The source term 3-form vanishes when the equations of motion are satisfied. 
For our specific case, the Noether charge 2-form can be expressed as 
\begin{align}
 \boldsymbol{Q}_{\boldsymbol{\xi}} \coloneqq \boldsymbol{Q}^{\mathrm{GR}}_{\boldsymbol{\xi}} + \boldsymbol{Q}^{\mathrm{EM}}_{\boldsymbol{\xi}},
\end{align}
where the components of the each term are given by 
\begin{align}
 \left(Q^{\mathrm{GR}}_{\boldsymbol{\xi}} \right)_{\mu\nu} \coloneqq - \frac{1}{16 \pi G} \epsilon_{\mu\nu\rho\sigma} \nabla^{\rho} \xi^{\sigma}, ~
 \left(Q^{\mathrm{EM}}_{\boldsymbol{\xi}} \right)_{\mu\nu} \coloneqq - \frac{1}{2} \epsilon_{\mu\nu\rho\sigma} F^{\rho\sigma} \xi^{\lambda} A_{\lambda}.
\end{align}
Here the components of 2-form can be understood as 
\begin{align}
 \boldsymbol{Q}_{\boldsymbol{\xi}}^{\mathrm{GR}} = \frac{1}{2} \left(Q_{\boldsymbol{\xi}}^{\mathrm{GR}} \right)_{\mu\nu} \mathbf{d}x^{\mu} \wedge \mathbf{d}x^{\nu},
\end{align}
and so on.
The source term 3-form $\boldsymbol{C}_{\boldsymbol{\xi}}$ can be evaluated as 
\begin{align}
\label{eq:source}
 (C_{\boldsymbol{\xi}})_{\mu\nu\rho} = \epsilon_{\sigma \mu\nu\rho} \left( T_{\lambda}{}^{\sigma}+ A_{\lambda} j^{\sigma} \right) \xi^{\lambda}.
\end{align}
Comparing Eq.~\eqref{defJ} and Eq.~\eqref{J_expressed_by_Q}, we obtain the Iyer--Wald identity
\begin{align}
 \mathbf{d}\boldsymbol{Q}_{\boldsymbol{\xi}} + \boldsymbol{C}_{\boldsymbol{\xi}} = \boldsymbol{\theta}\left(\phi; \mathsterling_{\boldsymbol{\xi}} \phi \right) - i_{\boldsymbol{\xi}} \boldsymbol{\mathcal{L}}(\phi).\label{IWidentity} 
\end{align}

By considering one parameter family of field configuration $\phi(\lambda)$ again, the $\lambda$ derivative of the identity \eqref{IWidentity} for $\phi(\lambda)$ can be expressed as
\begin{align}
 \mathbf{d} \left( \frac{d \boldsymbol{Q}_{\boldsymbol{\xi}}}{d \lambda} - i_{\boldsymbol{\xi}} \boldsymbol{\theta}\left(\phi; \frac{d \phi}{d\lambda}\right)  \right) 
 = \boldsymbol{\omega}\left(\phi; \frac{d \phi}{d \lambda}, \mathsterling_{\boldsymbol{\xi}} \phi \right)  - i_{\boldsymbol{\xi}} \left( \boldsymbol{E}(\phi) \cdot \frac{d \phi}{d \lambda}\right) 
- \frac{d \boldsymbol{C}_{\boldsymbol{\xi}}}{d \lambda}
.  \label{IWidentity_1st}
\end{align}
Here $\boldsymbol{\omega}$ is the pre-symplectic current 3-form defined by
\begin{align}
 \boldsymbol{\omega}\left(\phi; \frac{\partial \phi}{\partial \lambda_{1}}, \frac{\partial \phi}{\partial \lambda_{2}}\right)
\coloneqq \frac{\partial}{\partial \lambda_{1}} \boldsymbol{\theta}\left(\phi; \frac{\partial \phi}{\partial \lambda_{2}}\right) - \frac{\partial}{\partial \lambda_{2}} \boldsymbol{\theta}\left(\phi; \frac{\partial \phi}{\partial \lambda_{1}}\right),
\end{align}
for any 2-parameter family of field configurations $\phi(\lambda_{1}, \lambda_{2})$. For the Einstein--Maxwell theory with the cosmological constant, $\boldsymbol{\omega}$ is given by
\begin{align}
 \boldsymbol{\omega} \coloneqq \boldsymbol{\omega}^{\mathrm{GR}} + \boldsymbol{\omega}^{\mathrm{EM}}.
\end{align}
The contribution from the Einstein--Hilbert term can be expressed as 
\begin{align}
 \boldsymbol{\omega}^{\mathrm{GR}} = \star \left( \star \boldsymbol{\omega}^{\mathrm{GR}} \right),  \label{eq:omegaGR}
\end{align}
with
\begin{align}
 (\star \omega^{\mathrm{GR}})_{\mu} \coloneqq \frac{1}{16 \pi G} P_{\mu}{}^{\nu \rho \sigma \lambda \tau}
\left(
\frac{\partial g_{\nu \rho}}{\partial \lambda_{2}} \nabla_{\sigma} \frac{\partial g_{\lambda \tau}}{\partial \lambda_{1}} - \frac{\partial g_{\nu \rho}}{\partial \lambda_{1}} \nabla_{\sigma} \frac{\partial g_{\lambda \tau}}{\partial \lambda_{2}}
\right). \label{eq:omegaGRstar}
\end{align}
Here $\star$ represents the Hodge dual
\footnote{
Here we use the convention,
\begin{align*}
 (\star \omega)_{\mu} \coloneqq \frac{1}{3!} \omega_{\nu\rho\sigma} \epsilon^{\nu\rho\sigma}{}_{\mu}.
\end{align*}
With this convention, for a $p$-form $\boldsymbol{\alpha}$ in a $n$-dimensional Lorentzian manifold, $\star\star \boldsymbol{\alpha} = (-1)^{p(n-p)+1}\boldsymbol{\alpha}$.
}
, and the tensor $P^{\mu\nu\rho\sigma\lambda\tau}$ is given by 
\begin{align}
 P^{\mu \nu \rho \sigma \lambda \tau } \coloneqq g^{\mu  \lambda }g^{\tau  \nu }g^{\rho  \sigma } - \frac{1}{2} g^{\mu  \sigma } g^{\nu  \lambda } g^{\tau  \rho } - \frac{1}{2} g^{\mu  \nu } g^{\rho  \sigma } g^{\lambda  \tau } - \frac{1}{2} g^{\nu \rho } g^{\mu \lambda } g^{\tau \sigma } + \frac{1}{2} g^{\nu \rho } g^{\mu \sigma } g^{\lambda \tau }.
 \label{eq:P}
\end{align}
The contribution from the Maxwell term can be expressed as 
\begin{align}
 \omega^{\mathrm{EM}}_{\mu\nu\rho} \coloneqq
\frac{\partial}{\partial \lambda_{2}} \left(\epsilon_{\sigma \mu\nu\rho} F^{\sigma \lambda} \right) \frac{\partial A_{\lambda}}{\partial \lambda_{1}}
-
 \frac{\partial}{\partial \lambda_{1}} \left(\epsilon_{\sigma \mu\nu\rho} F^{\sigma \lambda} \right) \frac{\partial A_{\lambda}}{\partial \lambda_{2}}.
 \label{eq:omegaEM}
\end{align}

Setting $\lambda = 0$ in Eq.~\eqref{IWidentity_1st}, we obtain 
\begin{align}
  \mathbf{d} \left( \delta \boldsymbol{Q}_{\boldsymbol{\xi}} - i_{\boldsymbol{\xi}} \boldsymbol{\theta}\left(\bar{\phi}; \delta \phi\right)  \right) 
 = \boldsymbol{\omega}\left(\bar{\phi}; \delta \phi, \mathsterling_{\boldsymbol{\xi}} \bar{\phi} \right) - i_{\boldsymbol{\xi}} \left( \boldsymbol{E}(\bar{\phi}) \cdot \delta \phi \right)
- \delta \boldsymbol{C}_{\boldsymbol{\xi}}
, \label{IWidentity_1st_0}
\end{align}
where $\bar{\phi}$ refers to background fields. We also express the derivatives evaluated at $\lambda=0$ as
\begin{align}
    \delta \phi=\left. \frac{d\phi}{d\lambda}\right|_{\lambda=0},\quad  \delta^2 \phi= \left. \frac{d^2\phi}{d\lambda^2}\right|_{\lambda=0},
\end{align}
and so on. Thus, the one parameter family of field configuration can be expanded as  
\begin{align}
 \phi(\lambda) = \bar{\phi} + \delta \phi \lambda + \frac{1}{2} \delta^2 \phi \lambda^2 + {\cal O}(\lambda^3).
\end{align}
The role of $\lambda$ is to keep track of the order of perturbations. The parameter $\lambda$ always appears in the combinations with $\delta(\dots)$, such as $\lambda\delta(\dots)$ and $\lambda^2\delta^2(\dots)$.
Integrating the identity \eqref{IWidentity_1st_0} over a hypersurface $\Sigma$, we obtain
\begin{align}
 \int_{\partial \Sigma} \left( \delta \boldsymbol{Q}_{\boldsymbol{\xi}} - i_{\boldsymbol{\xi}} \boldsymbol{\theta}\left(\bar{\phi}; \delta \phi\right) \right) = \int_{\Sigma} \left[ \boldsymbol{\omega}\left(\bar{\phi}; \delta \phi, \mathsterling_{\boldsymbol{\xi}} \bar{\phi} \right)  - i_{\boldsymbol{\xi}} \left( \boldsymbol{E}(\bar{\phi}) \cdot \delta \phi \right) - \delta \boldsymbol{C}_{\boldsymbol{\xi}} \right]. \label{firstorder_id}
\end{align}
This is the key identity useful to discuss the cosmic censorship conjecture as well as the 1st law of the black hole thermodynamics.

Since the identity \eqref{IWidentity} holds for any order of perturbations, one can derive the identity for the second order perturbations as is discussed in Refs.~\cite{Hollands:2012sf, Sorce:2017dst}. Taking the derivative of Eq.~\eqref{IWidentity_1st} with respect to $\lambda$, we obtain
\begin{align}
 &\mathbf{d} \left( \frac{d^2 \boldsymbol{Q}_{\boldsymbol{\xi}}}{d \lambda^2} - i_{\boldsymbol{\xi}} \frac{d}{d \lambda} \boldsymbol{\theta}\left(\phi; \frac{d \phi}{d\lambda}\right)  \right)  \notag\\
 &= \boldsymbol{\omega}\left(\phi; \frac{d \phi}{d \lambda}, \mathsterling_{\boldsymbol{\xi}} \frac{d\phi}{d \lambda} \right) - i_{\boldsymbol{\xi}} \left( \frac{d \boldsymbol{E}(\phi)}{d\lambda}  \cdot \frac{d \phi}{d \lambda}\right)
 - \frac{d^2 \boldsymbol{C}_{\boldsymbol{\xi}}}{d \lambda^2}
- i_{\boldsymbol{\xi}} \left( \boldsymbol{E}(\phi)  \cdot \frac{d^2 \phi}{d \lambda^2}\right)
\notag\\
& \qquad
 + (\text{terms vanishing when $\mathsterling_{\boldsymbol{\xi}} \phi = 0$})
.  \label{IWidentity_2nd}
\end{align}
Setting $\lambda = 0$, we obtain
\begin{align}
  \mathbf{d} \left( \delta^2 \boldsymbol{Q}_{\boldsymbol{\xi}} - i_{\boldsymbol{\xi}} \delta \boldsymbol{\theta}  \right) 
 &= \boldsymbol{\omega}\left(\bar{\phi}; \delta \phi, \mathsterling_{\boldsymbol{\xi}} \delta \phi \right)  - i_{\boldsymbol{\xi}} \left( \delta \boldsymbol{E}  \cdot \delta \phi\right)
- \delta^2 \boldsymbol{C}_{\boldsymbol{\xi}}
 - i_{\boldsymbol{\xi}} \left( \boldsymbol{E}(\bar{\phi})  \cdot \delta^2 \phi\right)
\notag\\
& \qquad
 + (\text{terms vanishing when $\mathsterling_{\boldsymbol{\xi}} \bar{\phi} = 0$}).
\end{align}
Then, integrating it over $\Sigma$, we obtain
\begin{align}
   \int_{\partial \Sigma}\left[ \delta^2 \boldsymbol{Q}_{\boldsymbol{\xi}} - i_{\boldsymbol{\xi}} \delta \boldsymbol{\theta} \right]
 &= \mathcal{E}_\Sigma + \int_{\Sigma} \left[
 - i_{\boldsymbol{\xi}} \left( \delta \boldsymbol{E}  \cdot \delta \phi\right)
 - \delta^2 \boldsymbol{C}_{\boldsymbol{\xi}}- i_{\boldsymbol{\xi}} \left( \boldsymbol{E}(\bar{\phi})  \cdot \delta^2 \phi\right)
\right] \notag\\
& \qquad
 + (\text{terms vanishing when $\mathsterling_{\boldsymbol{\xi}} \bar{\phi} = 0$}),\label{secondorder_id}
\end{align} 
where the canonical energy $\mathcal{E}$ is defined by 
\begin{align}
    \mathcal{E}_\Sigma(\bar{\phi}; \delta\phi, \mathsterling_{\boldsymbol{\xi}}\delta \phi) \coloneqq  \int _\Sigma \omega (\bar{\phi}; \delta\phi, \mathsterling_{\boldsymbol{\xi}} \delta \phi),
\end{align}
which is bilinear form of perturbations.

The second order identity \eqref{secondorder_id} plays an important role to confirm the weak cosmic censorship for de Sitter black holes, as is the asymptotically flat case \cite{Sorce:2017dst}, as we will see below.

\section{Black Holes in de Sitter Spacetime}
\label{sec:BHindS}
In this section, we will define a class of spacetimes referred to as ``asymptotically de~Sitter spacetimes'', which we will focus on in this paper. Unlike asymptotically flat spacetimes, they do not have a naturally defined spatial infinity. In order to establish the first law for black holes in asymptotically de~Sitter spacetimes, we introduce a substitute for the spatial infinity which is denoted by $S_\infty$ here. We will also examine that the Reissner--Nordstr\"{o}m--de Sitter spacetime satisfies a certain fall-off condition, under which our definition of mass and another known definition, the Abbott--Deser mass, both of which are defined in the next section, coincide.

\subsection{Asymptotically de Sitter Spacetimes}
Throughout this paper, we focus on an asymptotically de Sitter spacetime that is defined under the presence of the flat chart $(\tau, \rho, \theta, \varphi)$ of the de Sitter spacetime in the asymptotic region where the metric can be expressed as 
\begin{align}
 g_{\mu\nu} = g^{\mathrm{dS}}_{\mu\nu} + h_{\mu\nu}.
\end{align}
Here $g^{\mathrm{dS}}_{\mu\nu}$ is the metric of the de Sitter space and the components in the coordinates $(\tau, \rho, \theta, \varphi)$ can be described as
\begin{align}
  g^{\mathrm{dS}}_{\mu\nu} \mathbf{d}x^{\mu} \mathbf{d}x^{\nu} =  - \mathbf{d} \tau^2 + a(\tau)^2 \left(\mathbf{d} \rho^2 + \rho^2 (\mathbf{d}\theta^2 + \sin^2 \theta \mathbf{d} \varphi^2) \right).
\end{align}
Here the scale factor $a(\tau)$ is given by $a(\tau) = \mathrm{e}^{H \tau}$ and the constant $H$ is the Hubble parameter that is related to the positive cosmological constant $\Lambda$ as $H = \sqrt{\Lambda/3}$. 
The deviation from the de Sitter space, $h_{\mu\nu}$, is assumed to disappear in the limit $a(\tau) \rho \rightarrow \infty$.
To ensure the finiteness of our and the Abbott--Deser's mass formulas defined below, we require
\begin{align}
h_{\tau\tau} &= c_1 \frac{G m}{H^2 a(\tau)^3 \rho^3} + \mathcal{O}\left(\frac{1}{a(\tau)^4 \rho^4}\right), \label{adS1}\\ 
h_{\tau \rho} &=  \left( c_2 \frac{G m}{H a(\tau)^2 \rho^2} + \mathcal{O}\left(\frac{1}{a(\tau)^3 \rho^3}\right) \right)a(\tau), \label{adS2} \\
h_{\rho\rho} &=  \left( c_3 \frac{G m}{a(\tau) \rho} + \mathcal{O}\left(\frac{1}{a(\tau)^2 \rho^2}\right) \right) a(\tau)^2, \label{adS3}\\
h_{\theta\theta} &= \frac{1}{\sin^2 \theta} h_{\varphi \varphi} = \left( c_4 \frac{G m}{H^2 a(\tau)^3 \rho^3} + \mathcal{O}\left(\frac{1}{a(\tau)^4 \rho^4} \right) \right) a(\tau)^2 \rho^2 \label{adS4}.
\end{align}
Here we introduced a mass parameter $m$ and dimensionless constants $c_{1}, c_{2}, c_{3},$ and $c_{4}$. We call $a(\tau) \rho \rightarrow \infty$ the spatial infinity $S_{\infty}$. Note that it is not necessary for the deviation $h_{\mu\nu}$ to be small except for the spatial infinity. Thus we do not need to treat $h_{\mu\nu}$ as perturbations. 

The background de Sitter spacetime  possesses the ``static'' Killing vector 
\begin{align}
 \boldsymbol{\xi} = \boldsymbol{\partial}_{\tau} - H \rho \boldsymbol{\partial}_{\rho}. 
\end{align}
The norm of this vector can be evaluated as 
\begin{align}
 g^{\mathrm{dS}}_{\mu\nu} \xi^{\mu} \xi^{\mu} = -1 + H^2 a(\tau)^2 \rho^2. 
\end{align}
It is timelike, null, and spacelike for $a(\tau) \rho < H^{-1}$, $a(\tau) \rho = H^{-1}$, and $a(\tau) \rho > H^{-1}$ respectively. We choose the sign of $\xi^{\mu}$ so that it is future directed for $a(\tau) \rho < H^{-1}$. 

\subsection{Reissner--Nordstr\"{o}m--de Sitter Spacetime}
\label{Sec2.2RNdS}
Let us consider the static, spherically symmetric solution of the Einstein--Maxwell equations with a positive cosmological constant $\Lambda > 0$.
Assuming the absence of the magnetic field, one can obtain Reissner--Nordstr\"{o}m--de Sitter spacetime with an electric charge as,  
\begin{align}
\label{staticRNdS}
 g_{\mu\nu} \mathbf{d}x^{\mu} \mathbf{d}x^{\nu} &= - f(r) \mathbf{d}t^2 + \frac{\mathbf{d}r^2}{f(r)} + r^2 (\mathbf{d} \theta^2 + \sin^2 \theta \mathbf{d}\varphi^2 ), \\
\label{gaugefield} 
 A_{\mu} \mathbf{d}x^{\mu}  &= - \Phi(r) \mathbf{d}t,
\end{align}
where the function $f(r)$ and the Coulomb potential are given by
\begin{align}
\label{staticcoodf}
 f(r) &= 1 - \frac{2 G m}{r} + \frac{G k Q^2}{r^2} - H^2 r^2, \\
 \label{potential}
 \Phi(r) &= k \frac{Q}{r}, 
\end{align}
where $k = (4\pi)^{-1}$ is the Coulomb constant in our unit.
The static killing vector $\boldsymbol{\xi}$ is given by
\begin{align}
 \boldsymbol{\xi} = \boldsymbol{\partial}_{t}.\label{xistatic} 
\end{align}
Since the norm of $\boldsymbol{\xi}$ can be calculated as 
\begin{align}
 g_{\mu\nu} \xi^{\mu} \xi^{\nu} = - f(r),\label{gxixi}
\end{align}
$\boldsymbol{\xi}$ is time like only when $f(r) > 0$.
The Killing horizons can be defined by the positive root of $f(r) = 0$. Let us denote the position of one of the Killing horizons as $r_{\mathrm{H}}$.
The surface gravity $\kappa$ on $r = r_{\mathrm{H}}$ associated with $\boldsymbol{\xi}$ is 
defined by $\xi^{\nu}\nabla_{\nu} \xi^{\mu} = \kappa \xi^{\mu}$ and evaluated for a Reissner--Nordstr\"{o}m--de Sitter spacetime as  
\begin{align}
    \kappa=\frac{1}{2}f'(r_{\mathrm{H}})=- \frac{G k Q^2 - G m r_{\mathrm{H}} + H^2 r_{\mathrm{H}}^4}{r_{\mathrm{H}}^3}. 
\end{align}

\begin{figure}[t]
 \centering
 \includegraphics[width=0.7\textwidth, clip, trim=0 160 0 150]{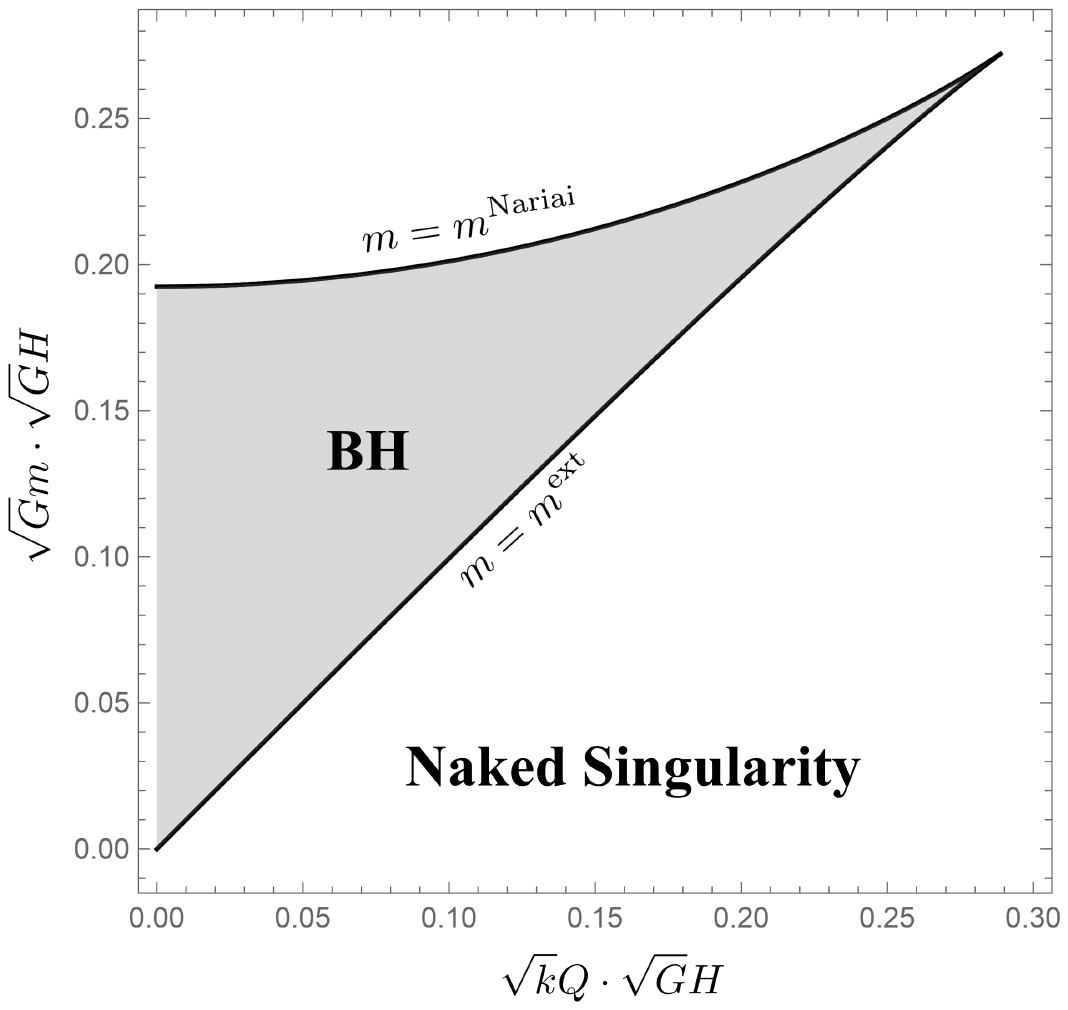}
 \caption{
The phase diagram for Reissner--Nordstr\"{o}m--de Sitter spacetimes:
The upper and lower curves correspond to $m = m^{\mathrm{Nariai}}$ and $m = m^{\mathrm{ext}}$, respectively. The gray region enclosed by these curves represents black hole spacetimes. The two curves intersect at $\sqrt{k} Q \cdot \sqrt{G} H = \sqrt{1/12} \approx 0.289$.
}
 \label{fig:spectrumofRNdS}
\end{figure}

Depending on the parameter $m$ and $Q$, a Reissner--Nordstr\"{o}m--de Sitter spacetime possesses at most three Killing horizons. 
First, let us focus on the case with a special choice of the parameters where the horizons degenerate, that is $f'(r_{\mathrm{H}}) = 0$. 
Thus the condition can be summarized as 
\begin{align}
 f(r_{\mathrm{H}}) &= 1 - \frac{2 G m}{r_{\mathrm{H}}} + \frac{G k Q^2}{r_{\mathrm{H}}^2} - H^2 r_{\mathrm{H}}^2 = 0,\\ 
 \left. \frac{d}{d r}  \left(r f(r)\right) \right|_{r = r_{\mathrm{H}}} &= 1 - \frac{G k Q^2}{r_{\mathrm{H}}^2} - 3 H^2 r_{\mathrm{H}}^2 = 0.
\end{align}
By solving these two equation, we obtain the two solutions.
One is given by 
\begin{align}
 r_{\mathrm{H}} &= r_{\mathrm{H}}^{\mathrm{ext}} \coloneqq \sqrt{\frac{1 - \sqrt{1 - 12 G k Q^2 H^2}}{6 H^2}}, \\
 m &= m^{\mathrm{ext}} \coloneqq \frac{1 + 12 G k Q^2 H^2 - \sqrt{1 - 12 G k Q^2 H^2}}{3 \sqrt{6} G H  \sqrt{1 - \sqrt{1 - 12 G k Q^2 H^2 }}}.
\end{align}
This case corresponds to the extremal black hole, where the inner and the outer black hole horizons coincide. 
The other solution is 
\begin{align}
 r_{\mathrm{H}} &= r_{\mathrm{H}}^{\mathrm{Nariai}} \coloneqq \sqrt{\frac{1 + \sqrt{1 - 12 G k Q^2 H^2}}{6 H^2}}, \\
 m &= m^{\mathrm{Nariai}} \coloneqq \frac{1 + 12 G k Q^2 H^2 + \sqrt{1 - 12 G k Q^2 H^2}}{3 \sqrt{6} G H  \sqrt{1 + \sqrt{1 - 12 G k Q^2 H^2 }}},
\end{align}
that corresponds to Nariai spacetime where the outer black hole horizon and the cosmological horizon coincide. 
When the mass parameter $m$ is between $m^{\mathrm{ext}}$ and $m^{\mathrm{Nariai}}$, thus, $ m^{\mathrm{ext}} < m < m^{\mathrm{Nariai}}$,  we have three positive roots of $f(r_{\mathrm{H}}) = 0$, which correspond to the inner black hole horizon, the outer black hole horizon, and the cosmological horizon. When the mass parameter $m$ is smaller than $m^{\mathrm{ext}}$ or greater than $m^{\mathrm{Nariai}}$, the spacetime singularity at the origin $r = 0$ is not hidden behind the event horizon. Hence the  phase diagram for the Reissner--Nordstr\"{o}m--de Sitter spacetime can be summarized as Fig.~\ref{fig:spectrumofRNdS}. 

The Reissner--Nordstr\"{o}m--de Sitter spacetime is asymptotic de Sitter space in the sense defined in the previous subsection.
To see it clearly, let us describe the spacetime in other coordinates. 
Let us introduce new time and radial coordinates $(\tau, \rho)$ by
\begin{align}
 t &= \tau - \zeta(\rho a(\tau)), \\
 r &= \rho a(\tau),
\end{align}
where $a(\tau) \coloneqq \mathrm{e}^{H \tau}$ and the function $\zeta$ is defined by 
\begin{align}
 \zeta(r) \coloneqq \int^{r} dr \left( \frac{1}{2 H r} + \sqrt{ \frac{1}{4 H^2 r^2} + \frac{1}{f(r)^2}} \right).
\end{align}
In this new coordinate, the Reissner--Nordstr\"{o}m--de Sitter metric can be expressed as
\begin{align}
 g_{\mu\nu} \mathbf{d}x^{\mu} \mathbf{d}x^{\nu}
& = - F( a(\tau) \rho) \mathbf{d} \tau^2 + \frac{1}{F(a(\tau) \rho)} a(\tau)^2 \mathbf{d} \rho^2 + a(\tau)^2 \rho^2 (\mathbf{d} \theta^2 + \sin^2 \theta \mathbf{d}\varphi^2 ), 
\end{align}
where the function $F(r)$ can be evaluated as
\begin{align}
F(r) = \frac{f(r)}{2} \left(1 - \sqrt{1 + \frac{4 H^2 r^2}{f(r)^2}}\right).
\end{align}
The static Killing vector \eqref{xistatic} in our coordinates can be expressed as :
\begin{align}
 \boldsymbol{\xi} = \boldsymbol{\partial}_{\tau} - H \rho \boldsymbol{\partial}_{\rho}, \label{xiinads}
\end{align}
which completely agree with the static Killing vector of the asymptotically de Sitter space. 

In the $r = a(\tau) \rho \rightarrow \infty$ limit, the function $F$ can be expanded as 
\begin{align}
F(r) = 1 - \frac{2 G M}{H^2 r^3} + \mathcal{O}\left(
\frac{1}{r^4}
\right),
\end{align}
Thus we obtain
\begin{align}
 g_{\mu\nu} = g^{\mathrm{dS}}_{\mu\nu} + h_{\mu\nu},
\end{align}
with 
\begin{align}
\label{httRNdS}
h_{\tau \tau} &= \frac{2 G m}{H^2 r^3} + \mathcal{O}\left( \frac{1}{r^4} \right),\\
\label{hrrRNdS}
h_{\rho\rho} &=  \mathcal{O}\left( \frac{1}{r^3} \right),
\end{align}
and the other components vanish. Thus, the coefficients $c_{i}$ in the asymptotic expansions \eqref{adS1} - \eqref{adS4} are read as $c_1 = 2,$ and $c_2 = c_{3} = c_{4} = 0$.

\section{First Law for de Sitter Black Holes}
\label{Sec:firstlaw}
In this section we apply the covariant phase space approach to asymptotically de Sitter spacetimes. 

\begin{figure}[t]
 \centering
\includegraphics[width=\textwidth]{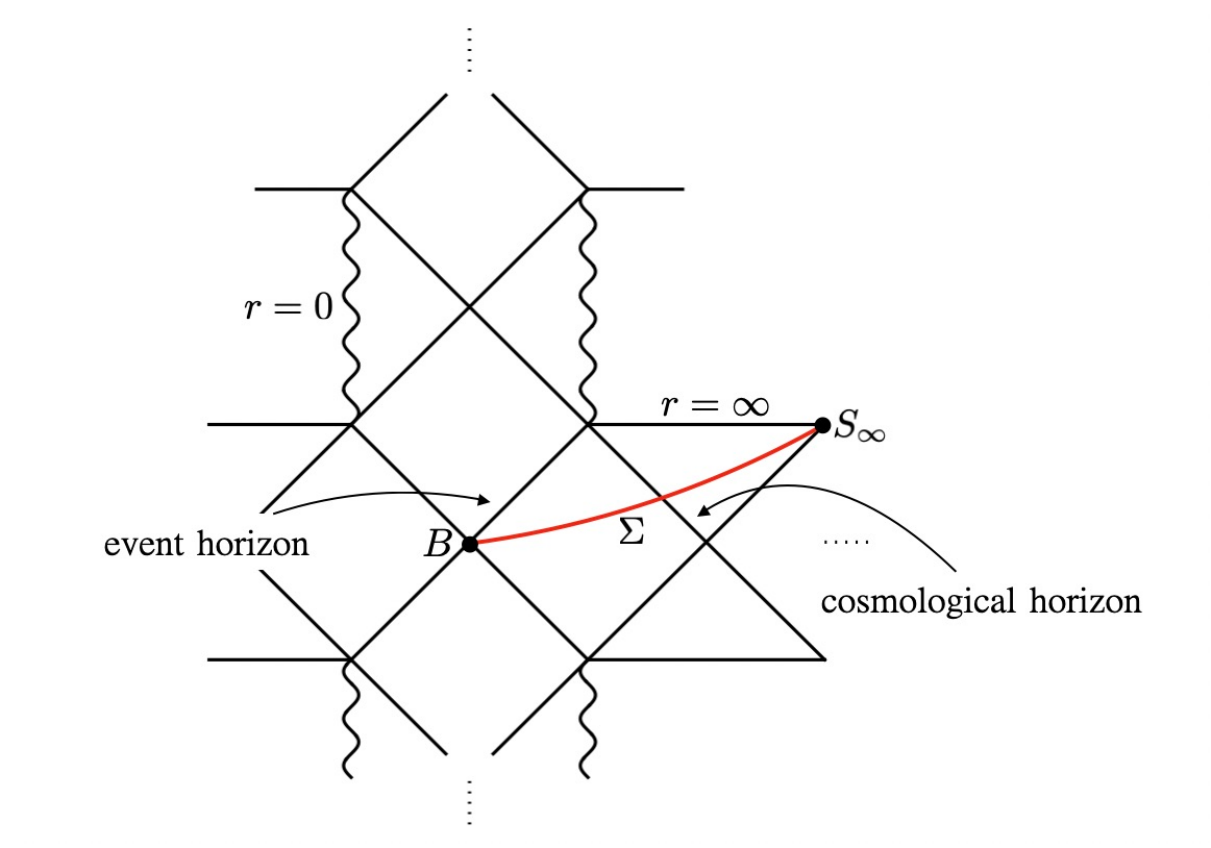}
 \caption{Penrose diagram of the back ground Reissner--Nordstr\"{o}m--de Sitter spacetime: We evaluate the Iyer-Wald identities on the hypersurface $\Sigma$ described by the red curve whose boundaries are the bifurcation surface $B$ and the spatial infinity $S_\infty$.}
 \label{RNdS_surface}
\end{figure}

Our setup is as follows:
we consider one parameter family of asymptotically de Sitter spacetimes $\{ g_{\mu\nu}(\lambda), A_{\mu}(\lambda) \}$ in the sense defined in Sec.~\ref{sec:BHindS}. Then, we assume that the background $\{ \bar{g}_{\mu\nu}, \bar{A}_{\mu} \} \coloneqq \{ g_{\mu\nu}(0), A_{\mu}(0) \}$ is Reissner--Nordstr\"{o}m--de Sitter solution \eqref{staticRNdS} and \eqref{gaugefield} with the parameters $\bar{m}$ and $\bar{Q}$. We note that the U(1) gauge condition for the background gauge field is fixed so that $\mathsterling_{\boldsymbol{\xi}} \bar{\boldsymbol{A}} = 0$. 
We will evaluate the identities \eqref{firstorder_id} and \eqref{secondorder_id} for the vector $\boldsymbol{\xi}$ and $\Sigma$ specified as follows:
for the vector $\boldsymbol{\xi}$, we use the static Killing vector of the background given by Eq.~\eqref{xiinads}. 
We note that there is a potential ambiguity of the overall normalization of the static Killing vector field. We will see that our choice, Eq.~\eqref{xiinads}, correctly reproduces the mass for the Schwarzschild--de~Sitter spacetime.
For the hypersurface $\Sigma$, we choose one which has the following two boundaries: the bifurcation surface denoted by $B$ and the spatial infinity denoted by $S_\infty$ shown in Fig.~\ref{RNdS_surface}. 
Hence the left hand sides of the identities \eqref{firstorder_id} and \eqref{secondorder_id} can be expressed as  
\begin{align}
\int_{\partial \Sigma} \left( \delta \boldsymbol{Q}_{\boldsymbol{\xi}} - i_{\boldsymbol{\xi}} \boldsymbol{\theta}\left(\bar{\phi}; \delta \phi\right) \right) =
\int_{S_{\infty}} \left( \delta \boldsymbol{Q}_{\boldsymbol{\xi}} - i_{\boldsymbol{\xi}} \boldsymbol{\theta}\left(\bar{\phi}; \delta \phi\right) \right)
- 
\int_{B} \left( \delta \boldsymbol{Q}_{\boldsymbol{\xi}} - i_{\boldsymbol{\xi}} \boldsymbol{\theta}\left(\bar{\phi}; \delta \phi\right) \right) ,
\label{lhsof1stidentity}
\end{align}
and
\begin{align}
\int_{\partial \Sigma} \left( \delta^2 \boldsymbol{Q}_{\boldsymbol{\xi}} - i_{\boldsymbol{\xi}} \delta \boldsymbol{\theta} \right) =
\int_{S_{\infty}} \left( \delta^2 \boldsymbol{Q}_{\boldsymbol{\xi}} - i_{\boldsymbol{\xi}} \delta \boldsymbol{\theta} \right)
- 
\int_{B} \left( \delta^2 \boldsymbol{Q}_{\boldsymbol{\xi}} - i_{\boldsymbol{\xi}} \delta \boldsymbol{\theta} \right).
\label{lhsof2ndidentity}
\end{align}

For the perturbations,
we fix the U(1) gauge degree of freedom around ${\cal H}$ by
\begin{align}
    \xi^\mu \delta A_\mu\overset{\mathcal{H}}{=} 0 .
\end{align}
In addition, we fix the diffeomorphism gauge degrees of freedom around ${\cal H}$ as follows: 
We impose that, near the horizon $\mathcal{H}$ of the background spacetime, each metric for any $\lambda$ can be written in the Gaussian null coordinates in the following form,
\begin{align}
    g_{\mu\nu}(\lambda)\mathbf{d}x^\mu \mathbf{d}x^\nu=2(\mathbf{d}v-v^2\alpha(\lambda) \mathbf{d}u - v\beta_A (\lambda) \mathbf{d}x^A)\mathbf{d}u+\gamma_{AB}(\lambda) \mathbf{d}x^A \mathbf{d}x^B,
\end{align}
where $x^A=(\theta,\varphi)$ are coordinates for $u, v$ constant 2-dimensional surfaces defined by $C(u,v)$ and $\alpha(\lambda)$, $\beta_{A}(\lambda)$, and  $\gamma_{AB}(\lambda)$ are one parameter family of functions of $x^{\mu}$, which are assumed to be finite at $v = 0$.
Note that we use a notation $\mathbf{d} v \mathbf{d} u = \frac{1}{2} \left(\mathbf{d} v \otimes \mathbf{d} u + \mathbf{d} u \otimes \mathbf{d} v \right)$ and so on.
We summarize the basic properties associated with the Gaussian null coordinates in \ref{App:GNC}.
We identify the Gaussian null coordinates for the background metric $\boldsymbol{g}(\lambda = 0)$ as given by Eq.~\eqref{eq:RNdSinGNC}.
In particular, the event horizon of the background spacetime ${\cal H}$ is identified as $v = 0$ with $u \geq 0$. Also, the bifurcation surface of the background $B$ is identified as a two-sphere $C(0,0)$ defined at $v = u = 0$.  
As derived in \ref{App:GNC}, in the Gaussian null coordinates, the background Killing vector $\boldsymbol{\xi} = \boldsymbol{\partial}_{t}$ can be expressed as 
\begin{align}
 \boldsymbol{\xi} = \kappa (u \boldsymbol{k} - v \boldsymbol{l}),
\end{align}
with the definitions $\boldsymbol{k} = \boldsymbol{\partial}_{u}$ and $\boldsymbol{l} = \boldsymbol{\partial}_{v}$.

Note that we will require additional assumptions in Sec.~\ref{sec:WCC} to discuss a gedanken experiment for the weak cosmic censorship conjecture.

\subsection{Contribution from the Spatial Infinity $S_{\infty}$: Mass Term}
At the spatial infinity, it is sufficient to consider the gravitational contribution since the electromagnetic one can be ignored because $F_{\mu\nu} \sim \mathcal{O}\left(\frac{1}{r^2}\right)$, $A_{\mu} \sim \mathcal{O}\left(\frac{1}{r}\right)$ when $r \rightarrow \infty$.
As is the ADM mass for asymptotically flat spacetimes, one can define the mass for asymptotically de Sitter space and there are several ways to define it. First of all, let us define the mass as Noether charge in Iyer and Wald's way. 

For defining a mass, we follows the definition in Refs.~\cite{Wald:1993nt,Iyer:1994ys,Iyer:1995kg} for asymptotically flat spacetimes, where the mass is defined through the surface integral on $S_{\infty}$ in the left-hand sides of the Iyer--Wald's identities, \eqref{lhsof1stidentity} and \eqref{lhsof2ndidentity}. Thus, we introduce the mass of asymptotically de Sitter black holes by
\begin{align}
\label{ourmass}
M(\phi(\lambda)) \coloneqq  \int_{S_{\infty}} \left( \boldsymbol{Q}_{\boldsymbol{\xi}}(\phi(\lambda)) - i_{\boldsymbol{\xi}} \boldsymbol{B}\left(\phi(\lambda) \right) \right) ,
\end{align}
where 2-form $\boldsymbol{B}$ is defined via
\begin{align}
    \frac{\partial}{\partial \zeta} \boldsymbol{B}(\phi(\lambda; \zeta)) = \boldsymbol{\theta}(\phi(\lambda; \zeta); \partial_{\zeta} \phi(\lambda; \zeta))\label{ourmassdash}
\end{align}
for any one parameter family of field configuration with $\zeta$.
Note that this one-parameter family with $\zeta$ needs not to coincide with the one with $\lambda$.

If one identifies $\zeta$ with $\lambda$, by taking a derivative along $\lambda$ and setting $\lambda = 0$, we obtain
\begin{align}
\delta M \coloneqq  \left. \frac{d}{d\lambda} M(\lambda) \right|_{\lambda = 0} = \int_{S_{\infty}} \left[ \delta \boldsymbol{Q}_{\boldsymbol{\xi}} - i_{\boldsymbol{\xi}} \boldsymbol{\theta}(\bar{\phi}; \delta \phi) \right] , \label{charge_infty}
\end{align}
and
\begin{align}
 \delta^2 M \coloneqq  \left. \frac{d^2}{d\lambda^2} M(\lambda) \right|_{\lambda = 0} = \int_{S_{\infty}} \left[ \delta^2 \boldsymbol{Q}_{\boldsymbol{\xi}} - i_{\boldsymbol{\xi}} \delta \boldsymbol{\theta} \right], \label{charge_infty_2nd}
\end{align}
which are the quantities appearing in Eqs.~\eqref{lhsof1stidentity} and \eqref{lhsof2ndidentity}.

Let us confirm that the mass $M(\lambda)$ defined by Eq.~\eqref{ourmass} actually can be interpreted as a mass. To confirm it, let us evaluate $M(\lambda)$ for asymptotically de Sitter class of metric \eqref{adS1} - \eqref{adS4},
\begin{align}
h_{\tau\tau}(\lambda) &= c_1(\lambda) \frac{G m(\lambda)}{H^2 a(\tau)^3 \rho^3},\\ 
h_{\tau \rho}(\lambda) &=  c_2(\lambda) \frac{G m(\lambda)}{H a(\tau)^2 \rho^2} a(\tau), \\
h_{\rho\rho}(\lambda) &= c_3(\lambda) \frac{G m(\lambda)}{a(\tau) \rho} a(\tau)^2,\\
h_{\theta\theta}(\lambda) &= \frac{1}{\sin^2 \theta} h_{\varphi \varphi}(\lambda) = c_4(\lambda) \frac{G m(\lambda)}{H^2 a(\tau)^3 \rho^3} a(\tau)^2 \rho^2.
\end{align}
Though we did not derive concrete expression for the 2-form $\boldsymbol{B}$, we can calculate the mass by introducing additional one parameter $\zeta$ as
$g_{\mu\nu}(\lambda; \zeta) = g^{\mathrm{dS}}_{\mu\nu} + \zeta h(\lambda)$. Here the parameter $\zeta$ is introduced so that $\phi(\lambda, 1) = \phi(\lambda)$ and $\phi(\lambda, 0)$ corresponds to the asymptotic massless de Sitter spacetime. 
By demanding that the mass of a de Sitter spacetime vanishes, we obtain  
\begin{align}
M(\lambda) = \int_{0}^{1} d \zeta  \int_{S_{\infty}} \left( \partial_{\zeta} \boldsymbol{Q}_{\boldsymbol{\xi}}(\phi(\lambda; \zeta)) - i_{\boldsymbol{\xi}} \boldsymbol{\theta}\left(\phi(\lambda;\zeta); \partial_{\zeta} \phi(\lambda,\zeta) \right) \right).\label{massintegral} 
\end{align}
Then, by evaluation the integral with our falloff conditions, we obtain,
\begin{align}
 M(\lambda) = \frac{1}{2} \left(c_{1}(\lambda) - 2 c_{2}(\lambda) + c_{3}(\lambda) - 3 c_{4}(\lambda) \right) m(\lambda).
 \label{eq:Mfalloff}
\end{align}
Additionally, if the metric $g_{\mu\nu}(\lambda)$ is the exact Reissner--Nordstr\"{o}m--de Sitter spacetime with a mass parameter $m(\lambda)$, we obtain $c_{1} = 2, c_{2} = c_{3} = c_{4} = 0$ and hence 
\begin{align}
 M(\lambda) = m(\lambda).\label{MRNdS}
\end{align}
Thus our mass reduces the usual mass parameter for Reissner--Nordstr\"{o}m--de Sitter spacetime. 
We should note that this is because the normalization ambiguity of the Killing vector is fixed by Eq.~\eqref{xiinads}.
Note that our falloff conditions ensures that the mass can be calculated from the first order perturbations in $\zeta$. If we relax the fall off conditions, for example, starting from $h_{\tau\tau}= {\cal O}(1/(\rho a(t)))$ and so on, the correct mass can be obtained if one calculates our mass formula up to ${\cal O}(\zeta^3)$.


Let us compare our definition of the mass in de Sitter spacetime as another known definition called Abbott--Deser mass ~\cite{Abbott:1981ff,Nakao:1994fj,Shiromizu:1993fi}.
The Abbott -- Deser mass $M^{\mathrm{AD}}$ is expressed by the surface integral of the Abbott--Deser charge 2-form $\boldsymbol{Q}^{\mathrm{AD}}$ at the spatial infinity in our sense:
\begin{align}
 M^{\mathrm{AD}} \coloneqq \int_{S_{\infty}} \boldsymbol{Q}^{\mathrm{AD}},\label{defMAD}
\end{align}
where the Abbott -- Deser charge 2-form can be defined by 
\begin{align}
 Q^{\mathrm{AD}}_{\mu\nu} \coloneqq \frac{1}{16 \pi G} \epsilon^{\mathrm{dS}}_{\mu\nu\rho\sigma} \left( - \nabla^{\mathrm{dS}}_{\beta}K^{\rho\sigma\alpha\beta} + K^{\rho\beta\alpha\sigma} \nabla^{\mathrm{dS}}_{\beta} \right) \left( g^{\mathrm{dS}}_{\alpha\gamma} \xi^{\gamma}\right) .\label{defQAD}
\end{align}
Here $\nabla^{\mathrm{dS}}$ and $\epsilon^{\mathrm{dS}}_{\mu\nu\rho\sigma}$ represent the covariant derivative and the volume form with respect to the background de Sitter space. 
The tensor $K^{\mu\alpha\nu\beta}$ is defined by
\begin{align}
 K^{\mu\alpha\nu\beta} \coloneqq \frac{1}{2} \left((g^{\mathrm{dS}})^{\mu\beta}H^{\nu\alpha} + (g^{\mathrm{dS}})^{\nu\alpha} H^{\mu\beta} - (g^{\mathrm{dS}})^{\mu\nu} H^{\alpha\beta} - (g^{\mathrm{dS}})^{\alpha\beta} H^{\mu\nu} \right),
\end{align}
with 
\begin{align}
 H^{\mu\nu} = \left((g^{\mathrm{dS}})^{\mu\rho} (g^{\mathrm{dS}})^{\nu\sigma} - \frac{1}{2} (g^{\mathrm{dS}})^{\mu\nu} (g^{\mathrm{dS}})^{\rho\sigma} \right) h_{\rho\sigma}.
\end{align}
Our formula \eqref{defMAD} corresponds to Eq.~(2.19) in the original paper~\cite{Abbott:1981ff}, though we express it in terms of the differential forms
\footnote{Our expression of the Abbott--Deser mass, Eq.~\eqref{defMAD} with Eq.~\eqref{defQAD}, looks $(-1)$ times the Eq.(2.19) in Ref.~\cite{Abbott:1981ff}.
However, these two are equivalent because we consider the future directed killing vector $\xi^{\mu}$, while the past directed one is used in Ref.~\cite{Abbott:1981ff}.
}
.

One can directly show that our mass coincides with the expression of Abbott--Deser mass
 at the linear order in $h_{\mu\nu}$.
Again, by introducing the one parameter $\zeta$ by $g_{\mu\nu}(\lambda) = g_{\mu\nu}^{\mathrm{dS}} + \zeta h_{\mu\nu}(\lambda)$, our mass can be expressed through \eqref{massintegral} as 
\begin{align}
 M(\lambda) = \int_{S_{\infty}} \left( \boldsymbol{Q}^{(1)}_{\boldsymbol{\xi}} - i_{\boldsymbol{\xi}} \boldsymbol{\theta}\left(\phi^{dS}; h \right) + {\cal O}(h^2)\right) ,
\end{align}
where $\boldsymbol{Q}^{(1)}_{\boldsymbol{\xi}} \coloneqq \partial_{\zeta} \boldsymbol{Q}_{\boldsymbol{\xi}}|_{\zeta = 0}$. The first term can be evaluated explicitly as 
\begin{align}
& \left[ Q^{(1)}_{\boldsymbol{\xi}} - i_{\boldsymbol{\xi}} \theta \left(\phi^{dS}; h \right)\right]_{\mu\nu} \notag\\
&  =
- \frac{1}{2} \epsilon^{\mathrm{dS}}_{\mu\nu}{}^{\rho\sigma}\left[\frac{1}{8\pi G}\left(\frac{1}{2} h \nabla^{\mathrm{dS}}_{\rho}\xi_{\sigma}-h_{\rho}{}^{\lambda} \nabla^{\mathrm{dS}}_\lambda \xi_{\sigma}+ \nabla^{\mathrm{dS}}_{\rho}h_{\sigma \lambda}\xi^\lambda - \nabla^{\mathrm{dS}}_{\lambda} h^{\lambda}{}_{\sigma}\xi_{\rho}+ \nabla^{\mathrm{dS}}_{\sigma}h\xi_{\rho}\right) \right],\label{IWmasslinearinh}
\end{align}
where in this expression the suffixes are lifted and lowered by $g^{\mathrm{dS}}_{\mu\nu}$ and its inverse: $h \coloneqq  h_{\mu\nu} g^{\mathrm{dS}}{}^{\nu\mu}$, $h_{\mu}{}^{\nu}\coloneqq h_{\mu\rho} g^{\mathrm{dS}}{}^{\rho\nu}$, and $\xi_{\mu} \coloneqq g^{\mathrm{dS}}_{\mu\nu} \xi^{\nu}$. 
One can check that the right hand side of Eq.~\eqref{IWmasslinearinh} agrees with the expression of the Abbott--Deser charge 2-form \eqref{defQAD}. Thus we find that 
our mass coincide with the Abbott--Deser mass linear in $h$:
\begin{align}
   M(\lambda) = M^{AD}(\lambda) + \mathcal{O}(h^2).
\end{align}
 Note that the Abbott--Deser mass, when evaluated under the falloff conditions, coincides with Eq.~\eqref{eq:Mfalloff}. Thus, under the falloff conditions, our mass coincides with the Abbott--Deser mass without any $\mathcal{O}(h^2)$  difference.
 If we relaxes the falloff conditions, the linear expression for the Abbott--Deser mass does not return the correct mass.

\subsection{Contribution from the Bifurcation Surface $B$: Area and Charge Terms}
Now let us evaluate the boundary contribution in the left hand sides of Eqs.~\eqref{lhsof1stidentity} and \eqref{lhsof2ndidentity} from the bifurcation surface $B$ of background black hole spacetime, described by
\begin{align}
\label{QonB}
    \int_{B}\left( \delta\boldsymbol{Q}_{\boldsymbol{\xi}} - i_{\boldsymbol{\xi}} \boldsymbol{\theta}\left(\bar{\phi};\delta\phi\right) \right) 
~
\text{and}
    \int_{B}\left( \delta^2\boldsymbol{Q}_{\boldsymbol{\xi}} - i_{\boldsymbol{\xi}} \delta \boldsymbol{\theta} \right) 
.
\end{align}
Although the analysis near the horizon follows the same approach as in the asymptotically flat case \cite{Wald:1993nt,Iyer:1994ys,Iyer:1995kg,Hollands:2012sf,Sorce:2017dst}. For completeness, we provide the detailed calculations here to ensure the paper is self-contained.

By definition of the bifurcation surface, the background Killing vector $\boldsymbol{\xi}$ vanishes on $B$, $ \boldsymbol{\xi}\overset{B}{=} 0$. Therefore, the second term of each integral in Eq.~\eqref{QonB} vanishes. Thus, we only need to evaluate the contributions from $\delta \boldsymbol{Q}_{\boldsymbol{\xi}}$ terms. 

Let us evaluate the gravitational contribution non-perturbatively. Thus, let us evaluate
\begin{align}
 \int_{B} \boldsymbol{Q}^{\mathrm{GR}}_{\boldsymbol{\xi}} = - \frac{1}{16 \pi G} \int_{B} \frac{1}{2} \epsilon_{\mu\nu\rho\sigma} \nabla^{[\rho} \xi^{\sigma]} \mathbf{d}x^{\mu} \wedge \mathbf{d}x^{\nu} \label{QGRB}.
\end{align}
The first observation is that, since $\xi^{\mu} \overset{B}{=} 0$, we obtain
\begin{align}
 \nabla_{\mu} \xi_{\nu} &= \bar{\nabla}_{\mu} \xi_{\nu} + \Gamma^{\rho}{}_{\mu\nu} \xi_{\rho} \overset{B}{=} \bar{\nabla}_{\mu} \xi_{\nu} \label{DxionB},
\end{align}
where $\Gamma^{\rho}{}_{\mu\nu}$ is defined by
\begin{align}
 \Gamma^{\rho}{}_{\mu\nu} \coloneqq \frac{1}{2} g^{\rho\sigma} \left( \bar{\nabla}_{\mu} g_{\sigma\nu} + \bar{\nabla}_{\nu} g_{\sigma\mu} - \bar{\nabla}_{\sigma} g_{\mu\nu} \right).
\end{align}
Thus, the covariant derivative $\nabla$ in the expression \eqref{QGRB} can be replaced with $\bar{\nabla}$. 

Since $\boldsymbol{\xi}$ is normal to the surface $\mathcal{H}$, the Frobenius's theorem indicates $\xi_{\left[\mu\right.} \bar{\nabla}_\nu\xi_{\left.\rho\right]}=0$. It follows that there is $\boldsymbol{\omega}$ on $\mathcal{H}$ satisfying \footnote{ In general, if $\boldsymbol{\alpha} \wedge \boldsymbol{\beta} = 0$ for a 1-form $\boldsymbol{\alpha} (\neq 0)$ and $p$-form $\boldsymbol{\beta}$, there exists a $(p - 1)$-form $\boldsymbol{\gamma}$ which satisfies $\boldsymbol{\beta} = \boldsymbol{\gamma} \wedge \boldsymbol{\alpha}$.} 
\begin{align}
\label{eq:mnablaxi}
    \bar{\nabla}_{[\mu} \xi_{\nu]} \overset{\mathcal{H}}{=} 2\omega_{\left[\mu\right.}\xi_{\left.\nu\right]}.
\end{align}
Let us introduce the null vector $\boldsymbol{n} \coloneqq - \frac{1}{\kappa u} \boldsymbol{l}$, which satisfy $g_{\mu\nu} n^{\mu} \xi^{\nu} = -1$, and express $\boldsymbol{\omega}$ as
\begin{align}
 \omega_{\mu} = \omega_{n} n_{\mu} + \tilde{\omega}_{\mu},
\end{align}
where $\omega_{n} \coloneqq - \xi^{\mu} \omega_{\mu}$ and $\xi^{\mu} \tilde{\omega}_{\mu} = 0$. Note that $n_{\mu} \mathbf{d}x^{\mu}$ is independent of $\lambda$ as checked in Appendix \ref{app:GNCpert}, and hence 
$n_{\mu} = g_{\mu\nu} n^{\nu} = \bar{g}_{\mu\nu} n^{\nu}$. 

Next, let us fix the coefficient $\omega_n$.
Since $\boldsymbol{\xi}$ is a Killing vector field of the background spacetime and $\xi_{\mu} = g_{\mu\nu} \xi^{\nu} \overset{\cal{H}}{=} \bar{g}_{\mu\nu} \xi^{\nu}$ from Eq.~\eqref{xibaronH}, it satisfies
\begin{align}
\bar{\nabla}_{\mu} \xi_{\nu} \overset{\cal{H}}{=} \bar{\nabla}_{[\mu} \xi_{\nu]} \overset{\cal{H}}{=} 2 \omega_{[\mu} \xi_{\nu]}.  
\end{align}
Contracting with $\xi^{\mu}$, we obtain
\begin{align}
 \xi^{\mu} \bar{\nabla}_{\mu} \xi_{\nu}  \overset{\cal{H}}{=} - \omega_{n} \xi_{\nu}.
\end{align}
By the definition of the surface gravity $\kappa$, $\xi^\mu \bar{\nabla}_\mu \xi_\nu = \kappa \xi_\nu$, we can find $\omega_n=-\kappa$.
Thus Eq.~\eqref{eq:mnablaxi} can be expressed as
\begin{align}
    \bar{\nabla}_{[\mu}\xi_{\nu]} \overset{\mathcal{H}}{=} - 2 \kappa n_{[\mu}\xi_{\nu]}+2\tilde{\omega}_{[\mu}\xi_{\nu]}
= 2 \kappa l_{[\mu} k_{\nu]} + 2\tilde{\omega}_{[\mu}\xi_{\nu]}.
\end{align}
Therefore,
\begin{align}
 \epsilon_{\mu\nu\rho\sigma} \nabla^{[\rho} \xi^{\sigma]} &\overset{\mathcal{H}}{=} 
- 12 k_{[\mu} l_{\nu} \epsilon^{(2)}_{\rho\sigma]} \left( 2 \kappa l^{\rho} k^{\sigma} + 2 \tilde{\omega}^{\rho} \xi^{\sigma} \right) \notag\\
& = - 2 \kappa \epsilon^{(2)}_{\mu\nu} + \kappa u k_{[\mu} \epsilon^{(2)}_{\nu]\rho} \tilde{\omega}^{\rho},
\end{align}
where we use Eq.~\eqref{eq:epsilon432} for the first equality.
Since the second term vanishes when pulled back into $\mathcal{H}$, as shown in Eq.~\eqref{eq:2formpullbackcomponent}, we finally find that the gravitational contribution at the bifurcation surface is described by
\begin{align}
 \int_{B} \boldsymbol{Q}^{\mathrm{GR}}_{\boldsymbol{\xi}} =       -\frac{1}{16 \pi G} \int_B \frac{1}{2} \epsilon _{\mu\nu \rho \sigma} \nabla^{[\rho}\xi^{\sigma]} \mathbf{d} x^{\mu} \wedge \mathbf{d} x^{\nu}
     &=\frac{\kappa }{8\pi G} A_B, \label{QGR=AB}
\end{align}
where we define the area of the bifurcation surface $B$ by 
\begin{align}
A_{B} \coloneqq \int_{B} \boldsymbol{\epsilon}^{(2)}.
\end{align}
Finally, by expanding \eqref{QGR=AB} with respect to $\lambda$,  we obtain
\begin{align}
    \int_{B} \delta\boldsymbol{Q}^{\mathrm{GR}}_{\boldsymbol{\xi}} &=  \frac{\kappa}{8\pi G}\delta A_B
\qquad \mathrm{and} \qquad
    \int_{B} \delta^2 \boldsymbol{Q}^{\mathrm{GR}}_{\boldsymbol{\xi}} =  \frac{\kappa}{8\pi G}\delta^2 A_B.
\end{align}
Therefore, the gravitational contributions of the surface integral on $B$ can be expressed by the area of $B$. We note that, although we focus on the Einstein-Hilbert action here, the covariant phase space formalism is applicable to theories with higher derivative terms. In that case, the Bekenstein--Hawking entropy, $S_{\mathrm{BH}} = k_{\mathrm{B}} A_{B}/4 G \hbar$, should be replaced with the Wald entropy, where $k_{\mathrm{B}}$ and $\hbar$ are the Boltzmann constant and the reduced Planck constant respectively.

Next, let us evaluate the electromagnetic contribution.
From the gauge condition $\boldsymbol{A}(\lambda)\overset{B}{=} \bar{\boldsymbol{A}}$, it becomes
\begin{align}
    \int_{B} \boldsymbol{Q}^{\mathrm{EM}}_{\boldsymbol{\xi}} &= - \int_B \star \boldsymbol{F}~\xi^\rho \bar{A}_\rho =  \Phi_{\mathrm{H}} Q_B,
\end{align} 
where $\Phi_{\mathrm{H}}$ is electric potential at the horizon defined by $\Phi_{\mathrm{H}} \overset{\cal H}{=} - \xi^\rho \bar{A}_\rho$ and 
$Q_{B}$ is the total electric flux on B defined by  
\begin{align}
    Q_B \coloneqq \int_B \star \boldsymbol{F}=\frac{1}{2}\int_B \frac{1}{2} F^{\rho\sigma}\epsilon_{\rho\sigma\mu\nu}\mathbf{d}x^\mu\wedge \mathbf{d}x^\nu,
\end{align}
which can be regarded as the total charge included by $B$ through the Gauss law.
Expanding this expression with respect to $\lambda$, we obtain
\begin{align}
 \int_{B} \delta \boldsymbol{Q}^{\mathrm{EM}}_{\boldsymbol{\xi}}  = \Phi_{\mathrm{H}} \delta Q_{B} \qquad \mathrm{and} \qquad \int_{B} \delta^2 \boldsymbol{Q}^{\mathrm{EM}}_{\boldsymbol{\xi}}  = \Phi_{\mathrm{H}} \delta^2 Q_{B}.
\end{align}

Combining the gravitational and the electromagnetic contributions, the surface integral over the bifurcation surface can be obtained as 
\begin{align}
\int_{B} \left( \delta \boldsymbol{Q}_{\boldsymbol{\xi}} - i_{\boldsymbol{\xi}} \boldsymbol{\theta}\left(\bar{\phi}, \delta \phi \right) \right)  = \frac{\kappa}{8 \pi G}\delta A_B + \Phi_{\mathrm{H}} \delta Q_{B}, \label{charge_B}
\end{align}
and
\begin{align}
\int_{B} \left( \delta^2 \boldsymbol{Q}_{\boldsymbol{\xi}} - i_{\boldsymbol{\xi}} \delta \boldsymbol{\theta}\left(\bar{\phi}, \delta \phi \right) \right)  = \frac{\kappa}{8 \pi G}\delta^2 A_B + \Phi_{\mathrm{H}} \delta^2 Q_{B} . \label{charge_B_2nd}
\end{align}

\subsection{First Law for de Sitter Black Holes}
Now we are ready to derive the first law for de Sitter black holes.
As we explained in the beginning of this section, we assume that the background is Reissner--Nordstr\"{o}m--de Sitter solution and $\boldsymbol{\xi}$ is the static Killing vector of it. We also fixed U(1) gauge condition for the background gauge field as $\mathsterling_{\boldsymbol{\xi}}\bar{\boldsymbol{A}}=0$. Hence, the background satisfies $\boldsymbol{E}(\bar{\phi}) = 0$ and $\mathsterling_{\boldsymbol{\xi}}\bar{\phi} = 0$. Under these assumptions, the first and the second terms in the right hand side of the first order identity~\eqref{firstorder_id} vanish.
Here we use the fact that the pre-symplectic current 3-form $\boldsymbol{\omega}(\bar{\phi}; \delta \phi, \mathsterling_{\boldsymbol{\xi}} \bar{\phi})$ is a bilinear form of $\delta \phi$ and $\mathsterling_{\boldsymbol{\xi}} \bar{\phi}$, and it vanishes under $\mathsterling_{\boldsymbol{\xi}} \bar{\phi}=0$.
By substituting the results for the left hand side of the identity, Eqs.~\eqref{charge_infty} and \eqref{charge_B} via Eq.~\eqref{lhsof1stidentity}, we obtain the first law of de Sitter black hole as
\begin{align}
    \delta M- \frac{\kappa}{8\pi G}\delta A_B -\Phi_{\mathrm{H}} \delta Q_{B} = -\int_{\Sigma} \delta \boldsymbol{C}_{\boldsymbol{\xi}} . \label{firstlaw}
\end{align}
The right hand side represents the contribution of the source terms for the first order perturbations, which plays an important role in the Sorce--Wald formalism \cite{Sorce:2017dst} as we will see in the next section. If one assumes that the perturbations satisfy the electro-vacuum equations of motion, thus $\delta \boldsymbol{E} = 0$, as is the Iyer--Wald's setup~\cite{Wald:1993nt,Iyer:1994ys,Iyer:1995kg}, the right hand side of Eq.~\eqref{firstlaw} vanishes. 
Our expression completely agrees with that for asymptotically flat black holes, though the definition of the mass is different.

One can also derive the similar identity for the second order perturbations from Eq.~\eqref{secondorder_id}. By using $\boldsymbol{E}(\bar{\phi}) = 0$ and $\mathsterling_{\boldsymbol{\xi}} \bar{\phi} = 0$, as well as Eqs.~\eqref{charge_infty_2nd} and \eqref{charge_B_2nd} via Eq.~\eqref{lhsof2ndidentity} for the left hand side, we obtain
\begin{align}
\delta^2 M- \frac{\kappa}{8\pi G}\delta^2 A_B -\Phi_{\mathrm{H}} \delta^2 Q_{B} = \mathcal{E}_\Sigma - \int_{\Sigma}
 i_{\boldsymbol{\xi}} \left( \delta \boldsymbol{E}  \cdot \delta \phi\right) 
-
\int_{\Sigma} 
 \delta^2 \boldsymbol{C}_{\boldsymbol{\xi}}
. \label{firstlaw_2nd}
\end{align}
This is the first law of de Sitter black holes at the second order perturbations including the backreaction from the first order perturbations.

\section{Weak Cosmic Censorship for de Sitter Black Holes}
\label{sec:WCC}
In this section, we consider a gedanken experiment involving the injection of charge and energy into Reissner--Nordstr\"{o}m--de Sitter black holes to test whether we can overcharge asymptotically de Sitter black holes or not. Since we have formulated the first laws \eqref{firstlaw} and \eqref{firstlaw_2nd} analogous to those in the asymptotically flat cases, we now apply the method developed by Sorce and Wald \cite{Sorce:2017dst} for the asymptotically flat black holes to the context of de Sitter black holes.

\begin{figure}[t]
 \centering
 \includegraphics[width=\textwidth]{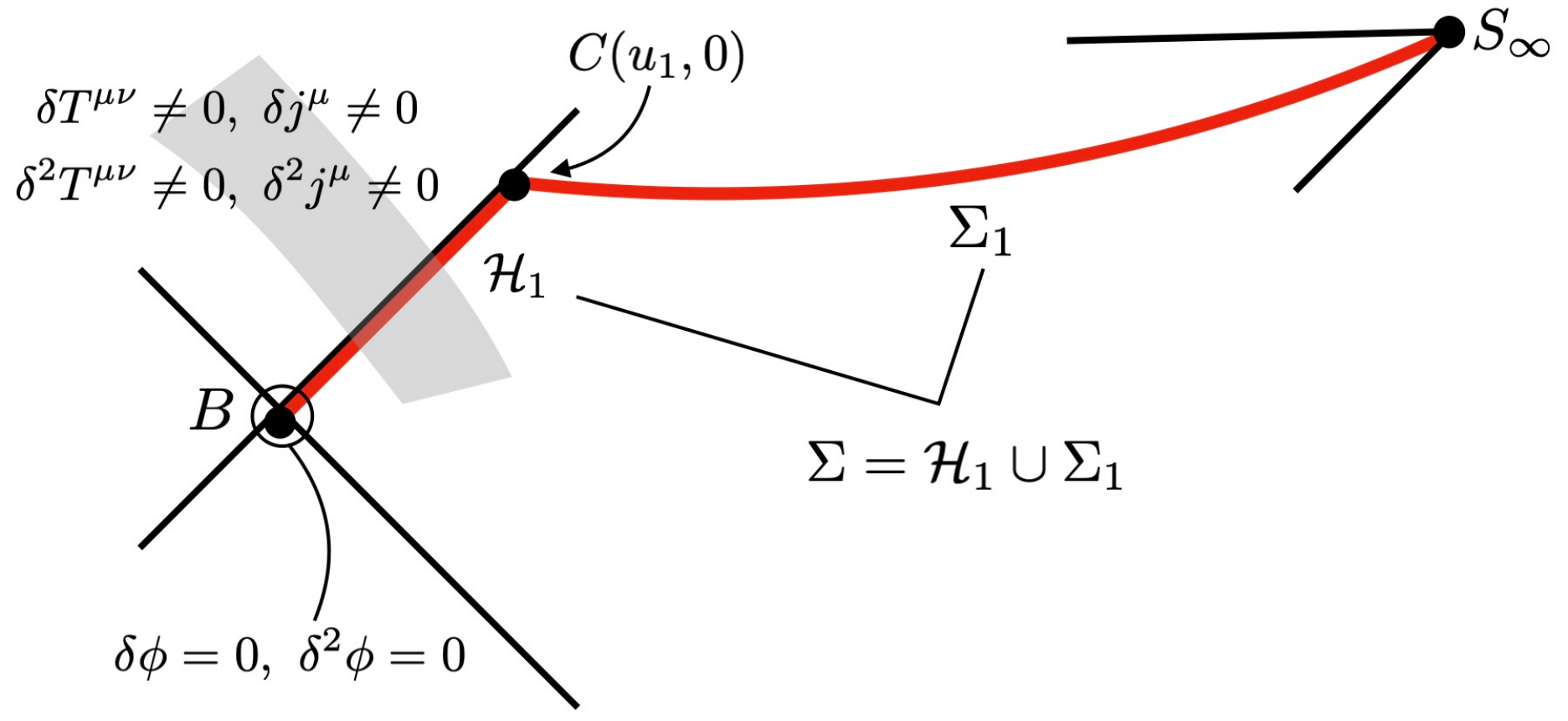}
 \caption{Setup to discuss a gedanken experiment for the weak cosmic censorship: The red curve means an example of a hypersurface $\Sigma$ that is the union of horizon part ${\cal H}_{1}$ and spacelike part $\Sigma_{1}$. The source term only exist on the gray region. }
 \label{surface}
\end{figure}

In addition to the setup in the previous section, we make additional assumptions as follows:
first, we assume the hypersurface $\Sigma$ is composed of two unions as is Fig.~\ref{surface}. Thus, we assume $\Sigma = {\cal H}_{1} \cup \Sigma_{1}$ where ${\cal H}_{1}$ is a portion of the background event horizon specified as $0 \leq u \leq u_{1}$ with $v = 0$, and $\Sigma_{1}$ is a spacelike hypersurface continued to the spatial infinity. 
The choice of the surface is almost same as Ref.~\cite{Sorce:2017dst}.
We assume that the source terms for the first and second order perturbations, $\delta T_{\mu\nu}, \delta j^{\mu}, \delta^2 T_{\mu\nu}, \delta^2 j^{\mu}$, are vanish except for the dark region in Fig.~\ref{surface}, in particular on $\Sigma_{1}$, which indicates that all the energies and charges are injected into background event horizon ${\cal H}_{1}$. Furthermore, we assume the boundary conditions for the perturbations such that $\delta \phi$ and $\delta^2 \phi$ vanish at the neighborhood of $B$, since our purpose is to investigate how the background spacetime is modified by the injection of the source. We additionally assume that the field configurations approach to Reissner--Nordstr\"{o}m--de Sitter solutions at the late time, especially on the $\Sigma_{1}$. As is demonstrated in Eq.~\eqref{MRNdS}, our mass $M(\lambda)$ can be understood as the mass parameter of such a final static spacetime $m(\lambda)$. 

\subsection{First Order Analysis}
Let us investigate the first order first law \eqref{firstlaw} in our setup.
Since we assumed that the perturbations vanish around $B$, the contributions from $B$ integral, $\delta A_B$ and $\delta Q_B$, vanish and hence the remaining part of the left hand side of Eq.~\eqref{firstlaw} becomes $\delta M$. Since the source term only contributes on ${\cal H}_{1}$, the right hand side of Eq.~\eqref{firstlaw} can be calculated as
\begin{align}
-\int_{\Sigma} \delta \boldsymbol{C}_{\boldsymbol{\xi}}&= -\int_{\Sigma}\delta \left( \frac{1}{3!} \epsilon_{\alpha \mu\nu\rho} \left( T_{\beta}{}^\alpha+ A_{\beta} j^{\alpha} \right) \xi^{\beta}\right)  \mathbf{d}x^{\mu} \wedge \mathbf{d}x^{\nu} \wedge  \mathbf{d}x^{\sigma}  \nonumber\\
 &= -\int_{\mathcal{H}_{1}}  \frac{1}{3!} \bar{\epsilon}_{\alpha\mu\nu\rho} \delta T_{\beta}{}^{\alpha} \xi^{\beta} \mathbf{d}x^{\mu} \wedge \mathbf{d}x^{\nu} \wedge  \mathbf{d}x^{\sigma} \notag\\
& \qquad - \int_{\mathcal{H}_{1}} \frac{1}{3!} \bar{\epsilon}_{\alpha \mu\nu\rho}  \xi^{\beta} \bar{A}_{\beta} \delta j^{\alpha} \mathbf{d}x^{\mu} \wedge \mathbf{d}x^{\nu} \wedge  \mathbf{d}x^{\sigma}, \label{eq:intdeltaC}
 \end{align}
 where, and hereafter, the suffixes are lifted and lowered by $\bar{g}_{\mu\nu}$ and its inverse. Here, we use the expression for $\delta \boldsymbol{C}_{\boldsymbol{\xi}}$ given by Eq.~\eqref{eq:source} in the first equality and use the fact that $\bar{T}_{\mu\nu} =  \bar{j}^\mu=0$ in the second equality.
 The first term can be calculated as, 
\begin{align}
     -\int_{\mathcal{H}_{1}}  \frac{1}{3!} \xi^{\beta} \bar{\epsilon}_{\alpha \mu\nu\rho} \delta T_{\beta}{}^{\alpha} \mathbf{d}x^{\mu} \wedge \mathbf{d}x^{\nu} \wedge  \mathbf{d}x^{\sigma} &= \int_{\mathcal{H}_{1}}  \delta T_{\mu\nu} \xi^\mu k^\nu \bar{\boldsymbol{\epsilon}}^{(3)} =\kappa \int_{\mathcal{H}_{1}} u \delta T_{\mu\nu} k^\mu k^\nu \bar{\boldsymbol{\epsilon}}^{(3)},
\end{align}
where we use Eq.~\eqref{eq:epsilon432} in the first equality and $\boldsymbol{\xi} \overset{\mathcal{H}}{=} \kappa u \boldsymbol{k}$ from Eq.~\eqref{eq:GNCkilling} in the second equality.
 Since $ - \xi^\mu \bar{A}_\mu \overset{\mathcal{H}}{=} \Phi_{\mathrm{H}}$ is constant, the second term in Eq.~\eqref{eq:intdeltaC} can be written as
 \begin{align}
- \int_{\mathcal{H}_{1}} \frac{1}{3!} \bar{\epsilon}_{\alpha \mu\nu\rho}  \xi^{\beta} \bar{A}_{\beta} \delta j^{\alpha} \mathbf{d}x^{\mu} \wedge \mathbf{d}x^{\nu} \wedge  \mathbf{d}x^{\sigma}
&= \Phi_{\mathrm{H}} \int_{\mathcal{H}_{1}} \star \delta \boldsymbol{j} = \Phi_{\mathrm{H}} \delta Q.  
\end{align}
Here we define $\delta Q \coloneqq \int_{\mathcal{H}_{1}} \star \delta \boldsymbol{j}$, which means the total flux of electromagnetic charge through the event horizon. Since there is no charge on $\Sigma_{1}$, $\delta Q$ can be understood as the first order perturbation of the total charge of this system.

Summarizing the calculations above, we obtain  
 \begin{align}
 \delta M - \Phi_{\mathrm{H}}\delta Q &= \kappa \int_{\mathcal{H}_{1}} u \delta T_{\mu\nu} k^\mu k^\nu \bar{\boldsymbol{\epsilon}}^{(3)}.
\end{align}
If we assume that the energy source thrown into the black hole satisfies the null energy condition, $\delta T_{\mu\nu} k^\mu k^\nu \geq 0$ ,  the relation between the change of the mass and the charge becomes
\begin{align}
    \delta M - \Phi_{\mathrm{H}} \delta Q \geq 0 \qquad (\text{under the null energy condition}) .
\end{align}
Thus, for a given amount of charge $\delta Q$, the minimum value for the increase in the total energy is
\begin{align}
    \delta M = \Phi_{\mathrm{H}} \delta Q.\label{optimalcondition}
\end{align}
Since this case corresponds to the situation where the extremal condition is most likely to be violated, in the following, we assume this optimal case.
Note that under the assumption of the null energy condition, this can be possible only when
\begin{align}
\left. \delta T_{\mu\nu} k^{\mu} k^{\nu} \right|_{\mathcal{H}_{1}} = 0.  \label{eq:optimalNEC}
\end{align}

\subsection{Second Order Analysis}
\label{sec:SW_2nd}
Let us evaluate the first low for the second order perturbations~\eqref{firstlaw_2nd}, with assuming the optimal situation \eqref{optimalcondition} for the first order perturbations.
Since we assumed that the second order perturbations also vanish at the neighborhood of the bifurcation surface $B$, the contributions from $B$ in the left-hand side of Eq.~\eqref{firstlaw_2nd} vanish, and hence, the left hand side of Eq.~\eqref{firstlaw_2nd} becomes $\delta^2 M$, as is the case of first order perturbations. 

Let us calculate the right hand side of Eq.~\eqref{firstlaw_2nd}.
The last term can be evaluated by the similar way to the first order perturbations. Noting that
$\left.\xi^\alpha \delta A_\alpha \right|_\mathcal{H}=0$ from the gauge condition and $\delta T_{\mu\nu} k^{\mu} k^{\mu} |_{\mathcal{H}_{1}} = 0$ from the optimal condition \eqref{eq:optimalNEC}, we obtain
\begin{align}
    -\int_{\mathcal{H}_{1}} \delta^2 \boldsymbol{C}_\xi&=-\int_{\mathcal{H}_{1}}  \delta^2 \left(\frac{1}{3!} \epsilon_{\alpha \mu\nu\rho} \left( T_{\beta}{}^{\alpha}+ j^{\alpha} A_{\beta}  \right)\xi^\beta \right) \mathbf{d}x^{\mu} \wedge \mathbf{d}x^{\nu} \wedge  \mathbf{d}x^{\sigma}  \nonumber \\
    &= -\int_{\mathcal{H}_{1}} \bar{\epsilon}_{\alpha \mu\nu\rho} \delta^2 T_{\beta}{}^{\alpha} \xi^\beta -\int_{\mathcal{H}_{1}} \delta^2 \left(\epsilon_{\alpha \mu\nu\rho}   j^{\alpha} \right) \bar{A}_{\beta}  \xi^\beta  \nonumber\\
    &= \kappa \int_{\mathcal{H}_{1}} u \delta^2 T_{\mu\nu} k^\mu k^\nu  \bar{\boldsymbol{\epsilon}}^{(3)} + \Phi_{\mathrm{H}} \delta^2 Q.
\end{align}
 
The second term of the right-hand side of Eq.~\eqref{firstlaw_2nd} vanishes identically, because 
the integral region can be replaced with $\mathcal{H}_{1}$ because $\delta \boldsymbol{E}$ vanishes on $\Sigma_{1}$, while the integral over $\mathcal{H}$ also vanishes because of $i_{\boldsymbol{k}} i_{\boldsymbol{\xi}} (\delta \boldsymbol{E} \cdot \delta \phi)$ = 0 because of the property of the pullback shown in Eq.~\eqref{inta=0}.

We evaluate the remaining part in the right hand side of Eq.~\eqref{firstlaw_2nd}, the canonical energy ${\cal E}_{\Sigma}$, which is consist of three contributions,
\begin{align}
     \mathcal{E}_\Sigma(\bar{\phi};\delta\phi)=\mathcal{E}^{\mathrm{GR}}_{\mathcal{H}_{1}}(\bar{\phi};\delta\phi)+\mathcal{E}^{\mathrm{EM}}_{\mathcal{H}_{1}}(\bar{\phi};\delta\phi)+\mathcal{E}_{\Sigma_1}(\bar{\phi};\delta\phi),
\end{align}
with
\begin{align}
    \mathcal{E}^{\mathrm{GR}}_{\mathcal{H}_{1}}(\bar{\phi};\delta\phi)&=\int_{\mathcal{H}_{1}} \boldsymbol{\omega}^{\mathrm{GR}}(\bar{\phi};\delta\phi,\mathsterling_{\boldsymbol{\xi}} \delta\phi) ~\\
    \mathcal{E}^{\mathrm{EM}}_{\mathcal{H}_{1}}(\bar{\phi};\delta\phi)&=\int_{\mathcal{H}_{1}} \boldsymbol{\omega}^{\mathrm{EM}}(\bar{\phi};\delta\phi,\mathsterling_{\boldsymbol{\xi}} \delta\phi) \\
    \mathcal{E}_{\Sigma_1}(\bar{\phi};\delta\phi)&= \int_{\Sigma_1} \boldsymbol{\omega}(\bar{\phi};\delta\phi,\mathsterling_{\boldsymbol{\xi}} \delta\phi).
\end{align}
 In the following, we will evaluate each of them individually. 

\subsubsection{$\mathcal{E}^{\mathrm{GR}}_{\mathcal{H}_{1}}$: Gravitational Contribution from the Horizon}
First, we focus on the gravitational contribution to the canonical energy from the horizon.
By using the expression of the pre-symplectic current 3-form \eqref{eq:omegaGR} and \eqref{eq:omegaGRstar}, it is expressed as 
\begin{align} 
\label{eq:canonical_GR_horizon}
    \mathcal{E}^{\mathrm{GR}}_{\mathcal{H}_{1}}(\bar{\phi};\delta\phi)&= \frac{1}{16\pi G} \int_{\mathcal{H}_{1}} \frac{1}{3!} \bar{P}^{\mu \nu \rho \sigma \lambda \tau}\left(\mathsterling_{\boldsymbol{\xi}}\delta g_{\nu\rho}\bar{\nabla}_{\sigma} \delta g_{\lambda\tau}-\delta g_{\nu\rho}\bar{\nabla}_\sigma \mathsterling_{\boldsymbol{\xi}}\delta g_{\lambda\tau}\right) \notag\\
& \qquad \qquad \qquad \qquad \qquad \qquad \times \bar{\epsilon}_{\mu\alpha\beta\gamma} \mathbf{d}x^\alpha\wedge \mathbf{d}x^\beta \wedge \mathbf{d}x^\gamma,
\end{align}
where $\bar{P}^{\mu \nu \rho \sigma \lambda \tau }$ is the background value $(\lambda = 0)$ of $P^{\mu\nu\rho\sigma\lambda\tau}$ defined by Eq.~\eqref{eq:P}.
Using Eq.~\eqref{eq:intVe}, it can be expressed as 
\begin{align}
    \mathcal{E}^{\mathrm{GR}}_{\mathcal{H}_{1}}(\bar{\phi};\delta\phi)&= \frac{1}{16\pi G} \int_{\mathcal{H}_{1}} \left[ - k_{\mu} \bar{P}^{\mu \nu \rho \sigma \lambda \tau}\left(\mathsterling_{\boldsymbol{\xi}}\delta g_{\nu\rho}\bar{\nabla}_{\sigma} \delta g_{\lambda\tau}-\delta g_{\nu\rho}\bar{\nabla}_\sigma \mathsterling_{\boldsymbol{\xi}}\delta g_{\lambda\tau}\right) \right] \bar{\boldsymbol{\epsilon}}^{(3)}. \label{eq:EGR2}
\end{align}
By using the properties, $ k^{\mu}\bar{\nabla}_{\mu} \delta g^\rho_{~\rho} \overset{\mathcal{H}}{=}0$, $\mathsterling_{\boldsymbol{\xi}} \delta g^\rho_{~\rho} \overset{\mathcal{H}}{=} 0$,  $k^\nu \mathsterling_{\boldsymbol{\xi}}\delta g_\nu^{~\sigma}\overset{\mathcal{H}}{=}0$, and $k^\lambda \bar{\nabla}_\sigma \mathsterling_{\boldsymbol{\xi}}\delta g^\sigma_\lambda\overset{\mathcal{H}}{=}0$, which correspond to Eqs.~\eqref{eq:Liektrdeltag}, \eqref{eq:kLiedeltag=0}, and \eqref{eq:kdelLiedeltag} shown in Appendix \ref{App:GNC}, 
the quantity in the square bracket in  Eq.~\eqref{eq:EGR2} can be simplified as 
\begin{align}
 \frac{1}{2} \mathsterling_{\boldsymbol{\xi}} \delta g^{\nu\rho} \left( k^{\sigma} \bar{\nabla}_{\sigma} \delta g_{\nu\rho} - 2 k^{\sigma} \bar{\nabla}_{\nu} \delta g_{\rho \sigma}\right)
- \frac{1}{2} \delta g^{\nu\rho} \left( 
k^{\sigma} \bar{\nabla}_{\sigma} \mathsterling_{\boldsymbol{\xi}} \delta g_{\nu\rho} - 2 k^{\sigma} \bar{\nabla}_{\nu} \mathsterling_{\boldsymbol{\xi}} \delta g_{\rho\sigma}
\right). \label{inthesquarebracket}
\end{align}
Using the properties $k^{\sigma} \delta g_{\rho\sigma} = 0$ and $k^{\sigma} \mathsterling_{\boldsymbol{\xi}} \delta g_{\rho\sigma}$ on $\mathcal{H}$, the covariant derivatives in the second terms within the brackets in Eq.~\eqref{inthesquarebracket} can be shifted to act on $k^{\sigma}$ and the quantities in the bracket can be expressed by the Lie derivatives with respect to $\boldsymbol{k}$,   
\begin{align}
&  \frac{1}{2} \mathsterling_{\boldsymbol{\xi}} \delta g^{\nu\rho} \left( k^{\sigma} \bar{\nabla}_{\sigma} \delta g_{\nu\rho} + 2 \bar{\nabla}_{\nu} k^{\sigma}  \delta g_{\rho \sigma}\right)
- \frac{1}{2} \delta g^{\nu\rho} \left( 
k^{\sigma} \bar{\nabla}_{\sigma} \mathsterling_{\boldsymbol{\xi}} \delta g_{\nu\rho} + 2 \bar{\nabla}_{\nu} k^{\sigma} \mathsterling_{\boldsymbol{\xi}} \delta g_{\rho\sigma}
\right) \notag\\
&\qquad =  \frac{1}{2} \mathsterling_{\boldsymbol{\xi}}\delta  g^{\nu\rho} \mathsterling_{\boldsymbol{k}} \delta g_{\nu\rho} - \frac{1}{2} \delta g^{\nu\rho}\mathsterling_{\boldsymbol{k}}\mathsterling_{\boldsymbol{\xi}}\delta g_{\nu\rho} \notag\\
&\qquad =
 \mathsterling_{\boldsymbol{\xi}}\delta  g^{\nu\rho} \mathsterling_{\boldsymbol{k}} \delta g_{\nu\rho} - \frac{1}{2}\mathsterling_{\boldsymbol{k}} (\delta g^{\nu\rho}\mathsterling_{\boldsymbol{\xi}}\delta g_{\nu\rho} )
 \label{inthesquarebracket2}
\end{align}
In addition, using the notion $\delta\sigma_{\mu\nu} = \frac{1}{2}\mathsterling_{\boldsymbol{k}} \delta g_{\mu\nu} = \frac{1}{2} \frac{1}{\kappa u} \mathsterling_{\boldsymbol{\xi}} \delta g_{\mu\nu}$
from Eq.~\eqref{deltasigma=}, Eq.~\eqref{inthesquarebracket2} can be expressed as 
\begin{align}
 4 \kappa u \delta \sigma^{\mu\nu}\delta \sigma_{\mu\nu} - \mathsterling_{\boldsymbol{k}} \left(\kappa u \delta g^{\mu\nu}\delta\sigma_{\mu\nu} \right).
\end{align}
Since the second term can be expressed as the total derivative,
\begin{align}
    \mathsterling_{\boldsymbol{k}} \left(\kappa u \delta g^{\mu\nu}\delta\sigma_{\mu\nu} \right)
    = k^{\rho} \bar{\nabla}_{\rho} \left( \kappa u \delta g^{\mu\nu}\delta\sigma_{\mu\nu} \right)
    =  \bar{\nabla}_{\rho} \left( k^{\rho} \kappa u \delta g^{\mu\nu}\delta\sigma_{\mu\nu} \right),
\end{align}
$\mathcal{H}_{1}$ integral of this term becomes boundary integral through Gauss's theorem \eqref{eq:Gauss}. Here, we use Eq.~\eqref{eq:barsigma} in the second equality.
Then, $\mathcal{E}^{\mathrm{GR}}_{\mathcal{H}}$ can be expressed as 
\begin{align}
 \mathcal{E}^{\mathrm{GR}}_{\mathcal{H}_{1}}(\bar{\phi}; \delta \phi) 
     &=\frac{\kappa}{4\pi G}\int_{\mathcal{H}_{1}} u \delta \sigma^{\mu\nu}\delta \sigma_{\mu\nu}\bar{\boldsymbol{\epsilon}}^{(3)}-\frac{\kappa}{16\pi G}\int_{\partial\mathcal{H}_{1}} u \delta g^{\mu\nu}\delta\sigma_{\mu\nu} \bar{\boldsymbol{\epsilon}}^{(2)} 
     \label{EGRwithbdr}.
\end{align}
Note that $\partial \mathcal{H}_{1} = B \cup C(u_{1}, 0)$.
Since we assumed that the perturbations are stationary at the surface $C(u_{1}, 0) = \mathcal{H} \cap \Sigma_{1}$ and vanish at the bifurcation surface $B$, the shear on $\partial \mathcal {H}_{1}$ vanishes,  $\delta \sigma_{\mu\nu}\overset{\partial\mathcal{H}_1}{=}0$. Hence the second term in Eq.~\eqref{EGRwithbdr} vanishes. 
Since $\delta \sigma_{\mu\nu}$ is a tensor projected on spacelike surfaces $C(u,0)$, we obtain
\begin{align}
    \mathcal{E}^{\mathrm{GR}}_{\mathcal{H}_{1}}(\bar{\phi}; \delta\phi)= \frac{\kappa}{4\pi G}\int_{\mathcal{H}_{1}} u \delta \sigma_{\mu\nu} \delta \sigma^{\mu\nu} \bar{\boldsymbol{\epsilon}}^{(3)} \geq 0,
\end{align}
where this term can be interpreted as the energy of the gravitational wave going into the background event horizon.

\subsubsection{$\mathcal{E}^{\mathrm{EM}}_{\mathcal{H}_{1}}$: Electromagnetic Contribution from the Horizon}
We now move on to the electromagnetic contribution to the horizon integration. The integration of the pre-symplectic current 3-form over $\mathcal{H}_{1}$ becomes
\begin{align}
\label{eq:canonicalenergy_EM_H}
    &\mathcal{E}^{\mathrm{EM}}_{\mathcal{H}_{1}}\left(\bar{\phi};\delta\phi\right)  \nonumber\\
    &\quad= \int_{\mathcal{H}_{1}} \frac{1}{3!} \bar{\epsilon}_{\sigma\mu\nu\rho}\left(\delta A_\lambda \mathsterling_{\boldsymbol{\xi}} \delta F^{\sigma\lambda}-\delta F^{\sigma\lambda}\mathsterling_{\boldsymbol{\xi}} \delta A_\lambda\right) \mathbf{d}x^\mu \wedge \mathbf{d}x^\nu \wedge \mathbf{d} x^\rho \nonumber\\
    &\qquad+\int_{\mathcal{H}_{1}}  \frac{1}{3!} \left(\left(\mathsterling_{\boldsymbol{\xi}} \delta \epsilon_{\sigma\mu\nu\rho}\right) \bar{F}^{\sigma\lambda}\delta A_\lambda -\left(\delta\epsilon_{\sigma\mu\nu\rho}\right) \bar{F}^{\sigma\lambda}\mathsterling_{\boldsymbol{\xi}}\delta A_\lambda\right) \mathbf{d}x^\mu \wedge \mathbf{d}x^\nu \wedge \mathbf{d} x^\rho. 
\end{align}
First, we focus on the last integral including $\bar{F}^{\sigma\lambda}$. From the assumption that the background spacetime is static, the flux of $\bar{T}^{\mathrm{EM}}_{\mu\nu}$ through the horizon vanishes. It reads that $\bar{T}^{\mathrm{EM}}_{\mu\nu}k^\mu k^\nu$ vanishes on $\mathcal{H}$ and thus implies that $\bar{T}^{\mathrm{EM}}_{\mu\nu}k^\mu$ is proportional to $k_\nu$ because of the dominant energy condition
\footnote{
If there are components tangent to the two surface $C(u,0)$, the norm of $\bar{T}^{\mathrm{EM}}_{\mu\nu}k^\mu$ becomes positive and this vector becomes spacelike, which leads to the violation of the dominant energy condition.
}
, which is satisfied for the energy momentum tensor of electromagnetic field.
Because of this, the form of $\bar{F}^{\sigma\lambda}$ is restricted as
\begin{align}
    \bar{F}^{\sigma\lambda} = \bar{v}^{[\sigma} k^{\lambda]} + \bar{w}^{\sigma\lambda},
\end{align}
where 
$\bar{v}^{\sigma}$ is a vector and $\bar{w}^{\sigma\lambda}$ is a tensor purely tangent to the horizon. Using this form and the gauge condition $k^\mu\delta A_\mu\overset{\mathcal{H}}{=}0$, the last integral in \eqref{eq:canonicalenergy_EM_H} vanishes.
The first integral can be expressed as
\begin{align}
& \int_{\mathcal{H}_{1}}  \frac{1}{3!} \bar{\epsilon}_{\sigma\mu\nu\rho}\left(\delta A_\lambda \mathsterling_{\boldsymbol{\xi}} \delta F^{\sigma\lambda}-\delta F^{\sigma\lambda}\mathsterling_{\boldsymbol{\xi}} \delta A_\lambda\right) \mathbf{d} x^{\mu} \wedge \mathbf{d} x^{\nu} \wedge \mathbf{d} x^{\rho}  \nonumber\\
    &\qquad =  \int_{\mathcal{H}_{1}}  \left[\mathsterling_{\boldsymbol{\xi}}\boldsymbol{\eta} - \frac{2}{3!}\bar{\epsilon}_{\sigma\mu\nu\rho}\delta F^{\sigma\lambda}\mathsterling_{\boldsymbol{\xi}} \delta A_\lambda \mathbf{d} x^{\mu} \wedge \mathbf{d} x^{\nu} \wedge \mathbf{d} x^{\rho} \right], \label{eq:canonicalenergy_EM_H_simple}
\end{align}
where we define 3-from $\boldsymbol{\eta}$ by
\begin{align}
 \boldsymbol{\eta} \coloneqq \frac{1}{3!} \bar{\epsilon}_{\sigma\mu\nu\rho}\delta A_\lambda  \delta F^{\sigma\lambda} \mathbf{d}x^\mu\wedge \mathbf{d}x^\nu \wedge \mathbf{d}x^\rho.
\end{align}
The Lie derivative of $\boldsymbol{\eta}$ can be expressed as 
\begin{align}
    \mathsterling_{\boldsymbol{\xi}} \boldsymbol{\eta} = i_{\boldsymbol{\xi}} \mathbf{d} \boldsymbol{\eta} + \mathbf{d}(i_{\boldsymbol{\xi}} \boldsymbol{\eta}).
\end{align}
The integral of the first term over $\mathcal{H}_{1}$ vanishes by Eq.~\eqref{inta=0}.  
Therefore, we obtain
\begin{align}
    \mathsterling_{\boldsymbol{\xi}} \boldsymbol{\eta} = \mathbf{d}(i_{\boldsymbol{\xi}} \boldsymbol{\eta}).
\end{align}
Thus, the first term of \eqref{eq:canonicalenergy_EM_H_simple} is only contributes on the surface $C(u_{1}, 0) = \mathcal{H} \cap \Sigma_{1}$. Since the perturbation is stationary on $C(u_{1}, 0)$, the electromagnetic flux vanishes. Following the same procedure as above, the form of $\delta F_{\mu\nu}$ is restricted as $\delta F_{\mu\nu} = \delta v_{[\mu} k_{\nu]} + \delta w_{\mu\nu}$, with a vector  
$\delta v_{\mu}$ and a tensor $\delta w_{\mu\nu}$ which is purely tangent to the horizon, and combining the gauge condition $k^\mu\delta A_\mu\overset{\mathcal{H}}{=}0$, the first term of \eqref{eq:canonicalenergy_EM_H_simple} also vanishes. The only left part is the second term of \eqref{eq:canonicalenergy_EM_H_simple}. A Lie-derivative of $\delta \boldsymbol{A}$ satisfies 
\begin{align}
\mathsterling_{\boldsymbol{\xi}} \delta \boldsymbol{A}=i_{\boldsymbol{\xi}} \mathbf{d} \left(\delta \boldsymbol{A}\right) + \mathbf{d} \left(i_{\boldsymbol{\xi}} \delta \boldsymbol{A}\right).\label{LiedeltaA}
\end{align}
Because of the gauge condition $ i_{\boldsymbol{\xi}}\delta \boldsymbol{A} =  \xi^\mu\delta A_\mu \overset{\cal H}{=}0$, the second term is normal to the horizon, which means it is proportional to $k_\lambda$. Since $\delta F^{\sigma\lambda}$ is anti symmetric, $\delta F^{\sigma\lambda}k_\lambda k_\sigma=0$, which reads that $\delta F^{\sigma\lambda} k_\lambda$, and hence $\delta F^{\sigma\lambda} (\mathbf{d} ~ i_{\boldsymbol{\xi}} \delta A)_{\lambda}$, is normal to $k^\sigma$.
By using Eq.~\eqref{eq:intVe} the contribution from the second term in Eq.~\eqref{LiedeltaA} vanishes in Eq.~\eqref{eq:canonicalenergy_EM_H_simple}. 
Then the final result of the canonical energy on ${\cal H}_{1}$ contributed from the electromagnetic part becomes
\begin{align}
    \mathcal{E}^{\mathrm{EM}}_{\mathcal{H}_{1}}\left(\bar{\phi};\delta\phi\right)&=- 2 \int_{\mathcal{H}_{1}} \frac{1}{3!} \bar{\epsilon}_{\sigma\mu\nu\rho}\xi^{\lambda}\delta F^{\sigma\rho}\delta F_{\lambda\rho} \mathbf{d}x^\mu \wedge \mathbf{d}x^\nu \wedge \mathbf{d}x^\rho.
\end{align}
Then, Using Eq.~\eqref{eq:intVe}, we obtain the expression, 
\begin{align}
    \mathcal{E}^{\mathrm{EM}}_{\mathcal{H}_{1}}\left(\bar{\phi};\delta\phi\right)
    &=
    2 \kappa \int_{\mathcal{H}_{1}}
    u 
    \bar{\gamma}^{\mu \nu}
    (\delta F_{\mu\sigma} k^{\sigma})(\delta F_{\nu\lambda} k^{\lambda})
      \bar{\boldsymbol{\epsilon}}^{(3)}
    \geq 0,
\end{align}
where we use $\bar{g}^{\mu\nu} \overset{\mathcal{H}}{=} k^{\mu} l^{\nu} + l^{\mu} k^{\nu} + \bar{\gamma}^{\mu\nu}$. 
\subsubsection{$\mathcal{E}_{\Sigma_1}$: Contribution from $\Sigma_1$}
Next, we move on to evaluating the contribution from the surface $\Sigma_1$, 
\begin{align}
    \mathcal{E}_{\Sigma_1}\left(\bar{\phi};\delta \phi\right) \coloneqq\int_{\Sigma_1}\boldsymbol{\omega}\left(\bar{\phi};\delta\phi,\mathsterling_{\boldsymbol{\xi}}\delta\phi\right).
\end{align} 
From the assumption, the spacetime settles into another Reissner--Nordstr\"{o}m--de Sitter solution in $\Sigma_1$. To emphasize it, we describe the perturbations $\delta \phi$ on $\Sigma_{1}$ as $\delta \phi^{\mathrm{RNdS}}$, and hence the canonical energy on $\Sigma_{1}$ can be expressed as
\begin{align}
    \mathcal{E}_{\Sigma_1} \left(\bar{\phi};\delta\phi\right) 
=\int_{\Sigma_1}\boldsymbol{\omega}\left(\bar{\phi};\delta\phi^{\mathrm{RNdS}},\mathsterling_{\boldsymbol{\xi}}\delta\phi^{\mathrm{RNdS}}\right).
\end{align}

Following the technique in Ref.~\cite{Sorce:2017dst}, let us evaluate this integral by considering another one parameter family of field configuration $\phi^{\mathrm{RNdS}}(\alpha)$, 
that is simply a family of Reissner--Nordstr\"{o}m--de Sitter solution with the mass and charge parameters given by
\begin{align}
    m^{\mathrm{RNdS}}(\alpha)&=\bar{M} + \alpha ~\delta M, \qquad  Q^{\mathrm{RNdS}}(\alpha)=\bar{Q} + \alpha~\delta Q. \label{staticperturbations}
\end{align}
Here $\bar{M}$, $\delta M$, $\bar{Q}$, and $\delta Q$ are chosen to be same as those of $\phi(\lambda)$.
Since there is no energy flux through $\mathcal{H}_{1}$ for static perturbations, there is no contributions to the canonical energy on $\mathcal{H}_{1}$ and we can freely extend the integral region from $\Sigma_1$ to $\Sigma$ for the static perturbations. Thus, the integral which we want to evaluated here is just a whole of the canonical energy over $\Sigma$ for the static perturbations, 
\begin{align}
\int_{\Sigma_1}\boldsymbol{\omega}\left(\bar{\phi};\delta\phi^{\mathrm{RNdS}},\mathsterling_{\boldsymbol{\xi}}\delta\phi^{\mathrm{RNdS}}\right)
=    \int_{\Sigma}\boldsymbol{\omega}\left(\bar{\phi};\delta\phi^{\mathrm{RNdS}},\mathsterling_{\boldsymbol{\xi}}\delta\phi^{\mathrm{RNdS}}\right) = {\cal E}(\bar{\phi}, \delta \phi^{\mathrm{RNdS}}).
\end{align}
Therefore we obtain
\begin{align}
 \mathcal{E}_{\Sigma_{1}}(\bar{\phi}; \delta \phi) = \mathcal{E}(\bar{\phi}; \delta \phi^{\mathrm{RNdS}})
\end{align}

Now let us evaluate the second order first law \eqref{firstlaw_2nd} for the static perturbations. Since there is no second order perturbations, $\delta^2 m^{\mathrm{RNdS}} = \delta^2Q^{\mathrm{RNdS}} = 0$, and $\delta\phi^{\mathrm{RNdS}}$ satisfies equations of motion, $\delta \boldsymbol{E}=\delta^2 \boldsymbol{C} =0$, the second order first law reduces to
\begin{align}
     - \frac{\kappa}{8 \pi G} \delta^2 A_B^{\mathrm{RNdS}}= \mathcal{E}(\bar{\phi};\delta\phi^{\mathrm{RNdS}}) .
\end{align}
Therefore, the canonical energy for the static perturbations $\mathcal{E}(\bar{\phi}, \delta \phi^{\mathrm{RNdS}})$, and hence the $\Sigma_{1}$ part of the canonical energy for the general perturbations $\mathcal{E}_{\Sigma_{1}}(\bar{\phi}; \delta \phi)$ can be expressed by the second order variation of the area of the bifurcation surface $B$ by the static perturbations \eqref{staticperturbations}.

Since the metric $\phi^{\mathrm{RNdS}}(\alpha)$ is explicitly given, the second order variation of the area $\delta^2 A^{\mathrm{RNdS}}$ can be evaluated straightforwardly. 
The result is 
\begin{align}
    &\delta^2 A_B^{\mathrm{RNdS}} = \frac{8 \pi  G \bar{r}_{\mathrm{H}}^2}{\left(G k \bar{Q}^2-G \bar{M} \bar{r}_{\mathrm{H}}+H^2 \bar{r}_{\mathrm{H}}^4\right)^3} \nonumber\\
&\qquad \qquad \qquad \times \Biggl[
    G(3 G \bar{M} - 2 \bar{r}_{\mathrm{H}}) \bar{r}_{\mathrm{H}}^3 \delta M^2
    + 2 G k \bar{Q} \bar{r}_{\mathrm{H}} (2 G k \bar{Q}^2 - 6 G \bar{M} \bar{r}_{\mathrm{H}}+ 3 \bar{r}_{\mathrm{H}}^2) \delta M \delta Q \notag\\
&\qquad \qquad \qquad \qquad      + k \bar{r}_{\mathrm{H}} (- 6 G \bar{M} \bar{r}_{\mathrm{H}}^2 + \bar{r}_{\mathrm{H}}^3 + 3 G^2 \bar{M} (- k \bar{Q}^2 + 3 \bar{M} \bar{r}_{\mathrm{H}}) ) \delta Q^2
    \Biggr].
\end{align}
Note that $\bar{r}_{\mathrm{H}}, \bar{M}, \bar{Q}$ and $H$ are not independent because these variables satisfy
\begin{align}
 1 - \frac{2 G \bar{M}}{\bar{r}_{\mathrm{H}}} + \frac{G k \bar{Q}^2}{\bar{r}_{\mathrm{H}}^2} - H^2 \bar{r}_{\mathrm{H}}^2 = 0.
\end{align} 
By using the assumption that the first order perturbations have optimally done, $\delta M = \Phi_{\mathrm{H}} \delta Q$, we obtain a simple expression, 
\begin{align}
 \delta^2 A_{B}^{\mathrm{RNdS}} = \frac{8 \pi G \bar{r}_{\mathrm{H}}^2}{G k \bar{Q}^2 - G \bar{M} \bar{r}_{\mathrm{H}} + H^2 \bar{r}_{\mathrm{H}}^4} k \delta Q^2.
\end{align}
Since the surface gravity $\kappa$ can be evaluated as
\begin{align}
\kappa = \frac{1}{2} \bar{f}'(\bar{r}_{\mathrm{H}}) = - \frac{G k \bar{Q}^2 - G \bar{M} \bar{r}_{\mathrm{H}} + H^2 \bar{r}_{\mathrm{H}}^4}{\bar{r}_{\mathrm{H}}^3},
\end{align} 
we obtain
\begin{align}
{\cal E}_{\Sigma_{1}}(\bar{\phi}; \delta \phi) = - \frac{\kappa}{8 \pi G} \delta^2 A_{B}^{\mathrm{RNdS}} = \frac{k}{\bar{r}_{\mathrm{H}}} \delta Q^2 .
\end{align}

\subsubsection{The Evaluation of the Identity for the Second Order Perturbations}
Now we have calculated all the terms in the second order fist law \eqref{firstlaw_2nd}, which now reduces to 
\begin{align}
\delta^2 M - \Phi_{\mathrm{H}} \delta^2 Q &=  \frac{k}{\bar{r}_{\mathrm{H}}} \delta Q^2 +
\kappa \int_{\mathcal{H}_{1}} u \delta^2 T_{\mu\nu}k^\mu k^\nu \bar{\boldsymbol{\epsilon}}^{(3)}\notag\\
&\quad  + \frac{\kappa}{4\pi G}\int_{\mathcal{H}_{1}} u \delta \sigma_{\mu\nu} \delta \sigma^{\mu\nu} \bar{\boldsymbol{\epsilon}}^{(3)}
+ 
2 \kappa  \int_{\mathcal{H}_{1}}
    u 
    \bar{\gamma}^{\mu \nu}
    (\delta F_{\mu\sigma} k^{\sigma})(\delta F_{\nu\lambda} k^{\lambda})
      \bar{\boldsymbol{\epsilon}}^{(3)}
\end{align}
Assuming that the non-electromagnetic energy-momentum tensor satisfies the null energy condition, the second term in the right hand side is positive. In addition, the third and fourth terms are automatically positive, and hence we obtain
\begin{align}
\label{eq:2nd_MQA}
    \delta^2 M- \Phi_{\mathrm{H}} \delta^2 Q \geq \frac{k}{\bar{r}_{\mathrm{H}}} \delta Q^2 \qquad (\text{under the null energy condition}).
\end{align}

\subsection{Comparison with the Extremal Condition}
From now on, we consider a gedanken experiment trying to overcharge a slightly non-extremal Reissner--Nordstr\"{o}m--de Sitter black hole.
Since our mass $M$ represents a mass parameter $m$ of a static Reissner--Nordstr\"{o}m-- de Sitter spacetime, which we assumed to be realized at late times, the extremal condition can be discussed with our mass $M$. 

Thus, we assume
\begin{align}
 \frac{\bar{M} - \bar{M}^{\mathrm{ext}}}{\bar{M}} =: \epsilon^2 \ll 1,
\end{align}
where $\bar{M}^{\mathrm{ext}}$ is defined by
\begin{align}
 \bar{M}^{\mathrm{ext}} \coloneqq \frac{1 + 12 G k \bar{Q}^2 H^2 - \sqrt{1 - 12 G k \bar{Q}^2 H^2}}{3 \sqrt{6} G H  \sqrt{1 - \sqrt{1 - 12 G k \bar{Q}^2 H^2 }}}.
\end{align}
If the perturbed spacetime is finally settled in a static black hole, the perturbed mass must be larger than the extremal value $M^{\mathrm{ext}}(\lambda)$, which is defined by 
\begin{align}
 M^{\mathrm{ext}}(\lambda) \coloneqq \frac{1 + 12 G k Q(\lambda)^2 H^2 - \sqrt{1 - 12 G k Q(\lambda)^2 H^2}}{3 \sqrt{6} G H  \sqrt{1 - \sqrt{1 - 12 G k Q(\lambda)^2 H^2 }}}.
\end{align}
Here the Taylor expansion of $M^{\mathrm{ext}}(\lambda)$ can be evaluated as 
\begin{align}
 M^{\mathrm{ext}}(\lambda) &=
\bar{M}^{\mathrm{ext}}
+ \frac{\sqrt{6} H k \bar{Q}}{\sqrt{1 - \sqrt{1 - 12 G k \bar{Q}^2 H^2}}} \delta Q \lambda \notag\\
& \qquad  + \frac{1}{2} \left( \frac{\sqrt{6} H k \bar{Q}}{\sqrt{1 - \sqrt{1 - 12 G k \bar{Q}^2 H^2}}}\delta^2 Q + \beta \delta Q^2\right) \lambda^2 + {\cal O}(\lambda^3),
\end{align}
where $\beta$ is given by
\begin{align}
 \beta \coloneqq - \frac{\sqrt{3}H}{\sqrt{2}} \frac{\sqrt{1 - \sqrt{1 - 12 G H^2 k \bar{Q}^2}}}{\sqrt{1 - 12 G H^2 k \bar{Q}^2}} k.
\end{align}
Since $\lambda$ appears in the combination with $\delta(\dots)$, $\mathcal{O}(\lambda)$ corresponds to linear combination of $\lambda\delta M$ and $\lambda \delta Q$. Thus $\mathcal{O}(\lambda^3)$ consists of linear combination of $\lambda^3\delta^3M$ and $(\lambda\delta M)^3$ and so on here.
Notice that the Coulomb potential on the horizon is given by
\begin{align}
 \Phi_{\mathrm{H}} = k \frac{\bar{Q}}{\bar{r}_{\mathrm{H}}}
 &=\frac{\sqrt{6} H k \bar{Q}}{\sqrt{1 - \sqrt{1 - 12 G k \bar{Q}^2 H^2}}} + \Delta \Phi_{\mathrm{H}} \epsilon + {\cal O}(\epsilon^2).
\end{align}

Here $\bar{r}_{\mathrm{H}}$ is expanded as $\bar{r}_{\mathrm{H}} = \bar{r}^{\mathrm{ext}}_{\mathrm{H}} + \epsilon \Delta \bar{r}_{\mathrm{H}} + \mathcal{O}(\epsilon^2)$, where $\bar{r}_{\mathrm{H}}^{\mathrm{ext}}$ and $\Delta \bar{r}_{\mathrm{H}}$ are given by
\begin{align}
\label{eq:rHepsilon}
   \bar{r}_{\mathrm{H}}^{\mathrm{ext}}&= \sqrt{\frac{1-\sqrt{1-12 G k Q^2 H^2}}{6H^2}}, \nonumber\\
   \Delta \bar{r}_{\mathrm{H}}&
   = \frac{1}{3 H}\sqrt{ - 1 + \frac{1 + 12 G H^2 k \bar{Q}^2}{\sqrt{1 - 12 G H^2 k \bar{Q}^2}}},
\end{align}
and
$\Delta \Phi_{\mathrm{H}}$ is defined by
\begin{align}
 \Delta \Phi_{\mathrm{H}} \coloneqq-k \frac{\bar{Q}}{(\bar{r}_{\mathrm{H}}^{\mathrm{ext}})^2}\Delta\bar{r}_{\mathrm{H}}=  - \frac{2 H k \bar{Q}}{1 - \sqrt{1 - 12 G H^2 k \bar{Q}^2}} \sqrt{ - 1 + \frac{1 + 12 G H^2 k \bar{Q}^2}{\sqrt{1 - 12 G H^2 k \bar{Q}^2}}}.
\end{align}
Then with this Coulomb potential the extremal mass can be expressed as 
\begin{align}
 M^{\mathrm{ext}}(\lambda) &=
\bar{M}^{\mathrm{ext}}
+ \Phi_{\mathrm{H}} \delta Q \lambda - \Delta \Phi_{\mathrm{H}} \delta Q \epsilon \lambda 
\notag\\ 
& \qquad + \frac{1}{2} \left( \Phi_{\mathrm{H}} \delta^2 Q + \beta \delta Q^2\right) \lambda^2
+ {\cal O}(\lambda^3, \epsilon^2 \lambda, \epsilon \lambda^2) .
\end{align}

Let us evaluate the difference between the actual mass $M(\lambda)$ and the extremal mass $M^{\mathrm{ext}}(\lambda)$,
\begin{align}
 F(\lambda) \coloneqq \frac{M(\lambda) - M^{\mathrm{ext}}(\lambda)}{\bar{M}}.
\end{align}
Then the function $F(\lambda)$ can be expanded as
\begin{align}
 F(\lambda) &= \epsilon^2 + \frac{1}{\bar{M}}\left(\delta M - \Phi_{\mathrm{H}} \delta Q \right) \lambda + \frac{\Delta \Phi_{\mathrm{H}}}{\bar{M}^{\mathrm{ext}}} \delta Q \epsilon \lambda
\notag \\ 
& \qquad   + \frac{1}{2 \bar{M}^{\mathrm{ext}}} \left( \delta^2 M - \Phi_{\mathrm{H}} \delta^2 Q - \beta \delta Q^2 \right) \lambda^2 + {\cal O}(\lambda^3, \epsilon^2 \lambda, \epsilon \lambda^2)
\label{expressionF}
\end{align}
We should note that we treat $\epsilon$ and $\lambda$ independently. For example, ${\cal O}(\epsilon^2 \lambda)$ is a contribution of $\epsilon^2\lambda\delta(\dots)$ which we assume to be small here.
As is the previous section, let us assume the first order perturbations have done optimally, $\delta M - \Phi_{\mathrm{H}} \delta Q = 0$. Then, by assuming the null energy condition for the second order perturbations, we can use the inequality \eqref{eq:2nd_MQA}. under these assumptions, the function $F(\lambda)$ can be evaluated as
\begin{align}
 F(\lambda) \geq \epsilon^2 + \frac{\Delta \Phi_{\mathrm{H}}}{\bar{M}^{\mathrm{ext}}} \delta Q \epsilon \lambda 
+ \frac{1}{2 \bar{M}^{\mathrm{ext}}} \left( \frac{k}{\bar{r}_{\mathrm{H}}} - \beta \right) \delta Q^2 \lambda^2 + {\cal O}(\lambda^3, \epsilon^2 \lambda, \epsilon \lambda^2)
\end{align}
By using the expressions, one can show that 
\begin{align}
 \frac{k}{\bar{r}_{\mathrm{H}}} - \beta  = \frac{1}{ 2 \bar{M}^{\mathrm{ext}}} \Delta \Phi^2.
\end{align}
Hence we obtain
\begin{align}
 F(\lambda) & \geq \epsilon^2 + 2 \frac{\Delta \Phi_{\mathrm{H}}}{2 \bar{M}^{\mathrm{ext}}} \delta Q \epsilon \lambda + \left( \frac{\Delta \Phi}{2 \bar{M}^{\mathrm{ext}}}  \delta Q \lambda\right)^2 + {\cal O}(\lambda^3, \epsilon^2 \lambda, \epsilon \lambda^2) \notag\\
&=  \left( \epsilon + \frac{\Delta \Phi}{2 \bar{M}^{\mathrm{ext}}} \delta Q \lambda \right)^2 + {\cal O}(\lambda^3, \epsilon^2 \lambda, \epsilon \lambda^2)
\label{positiveF}
\end{align}
Thus $F\geq0$ and the violation of the weak cosmic censorship conjecture due to the overcharging process does not occur as is the case of asymptotically flat black hole.

\section{Summary and Discussion}
In this paper, we formulate the first law for asymptotically de Sitter spacetimes based on the covariant phase space formalism. We focus on general asymptotically de Sitter perturbations around a Reissner--Nordstr\"{o}m--de Sitter black hole background in the Einstein--Maxwell system with a positive cosmological constant. The first laws are given by Eqs.~\eqref{firstlaw} and \eqref{firstlaw_2nd} for the first and the second order perturbations, respectively. Although our expressions have the same forms as in the asymptotically flat cases, they are written in terms of the Abbott--Deser mass rather than the ADM mass. 
Additionally, following the discussion by Sorce and Wald for asymptotically flat spacetimes in Ref.~\cite{Sorce:2017dst}, we apply our first laws to a thought experiment that involves overcharging a black hole by injecting energy and charge sources. By comparing the perturbed mass with the extremal mass, as shown in Eq.~\eqref{positiveF}, we confirm that Reissner--Nordstr\"{o}m de Sitter black holes cannot be overcharged, provided that the null energy conditions are satisfied. Therefore, our result supports the weak cosmic censorship conjecture in the context of the overcharging processes even in asymptotically de Sitter case.

In this paper, we focus only on the simple case of static background solutions in the Einstein--Maxwell system.
Therefore, extending our analysis beyond this setup is an important direction for future work.
One possible extension is to consider stationary backgrounds, specifically the Kerr--Newman--de Sitter class of background. Generalizing the first law in this way expected to be performed straightforwardly just by replacing the static Killing vector $\boldsymbol{\xi}$ to the killing vector associated with the horizon generators, as in the flat case~\cite{Wald:1993nt,Iyer:1994ys,Iyer:1995kg}. The expected results would including additional terms, $- \Omega_{\mathrm{H}} \delta J$ and $- \Omega_{\mathrm{H}} \delta^2 J$, in the first laws \eqref{firstlaw} and \eqref{firstlaw_2nd}. Here $\Omega_{\mathrm{H}}$ is the angular velocity of the background black hole, and $\delta J$ and $\delta^2 J$ are the first and second order perturbations of the Komar angular momentum, respectively. 
Similarly, the extension of the first law to more general system beyond the Einstein--Maxwell system is also straightforward.
In this case, the Bekenstein--Hawking entropy, thus the area term, in the first laws \eqref{firstlaw} and \eqref{firstlaw_2nd} would be replaced with the Wald entropy. However, the validity of the weak cosmic censorship conjecture when including higher derivative corrections remains uncertain even in the asymptotically flat case. For example, in Ref.~ \cite{Lin:2022ndf,Wang:2022umx}, it is pointed out that the null energy condition is insufficient to valid the weak cosmic censorship.
As a guideline for the condition for the validity of the weak censorship conjecture alternative to the null energy condition, requirement for the second law for the stationary perturbations is also discussed~\cite{Lin:2022ndf, Lin:2024deg, Wu:2024ucf}. 
In the asymptotically de Sitter case, it has been pointed out that the overcharging process is not prevented by the requirement of the second law for the total entropy of both the black hole and cosmological horizons.
In Appendix \ref{App:2nd}, we confirm that the requirement of the null energy conditions in Sorce--Wald formalism corresponds to that of the second law of the Bekenstein--Hawking entropy of 
only the black hole horizon. Thus, our results can be rephrased that the de Sitter black hole cannot be overcharged when the second law for black hole entropy holds. 

We would also like to comment on other processes that might lead to the creation of naked singularities in asymptotically de Sitter case.
As reviewed in Sec.~\ref{Sec2.2RNdS}, the singularity in the Reissner--Nordstr\"{o}m--de Sitter class of solution can be hidden only when $m^{\mathrm{ext}} \leq m \leq m^{\mathrm{Nariai}}$. In this paper, we confirmed that the first inequality, $m^{\mathrm{ext}} \leq m$, holds for the injection of energy and charge. However, our analysis does not rule out the violation of the second inequality, $m \leq m^{\mathrm{Nariai}}$. 
For instance, by using the first laws, one can derive the expression for $(m(\lambda) - m^{\mathrm{Nariai}}(\lambda))/\bar{m}$ similar to Eq.~\eqref{expressionF}, for the nearly Nariai background $ (\bar{m} - \bar{m}^{\mathrm{Nariai}})/\bar{m} = - \epsilon^2$.
However, since the first laws provide only lower bounds for $\delta m$ and $\delta^2 m$, we can only determine the lower bound for $(m(\lambda) - m^{\mathrm{Nariai}}(\lambda))/\bar{m}$. Thus, there is no guarantee that it remains non-positive. In fact, non-perturbative examples by injecting spherical energy sources \cite{Senovilla:2022bsn} indicate that the condition $m < m^{\mathrm{Nariai}}$ can be violated, although the resulting naked singularity, at least in the chargeless case, resembles a cosmological big crunch singularity, which appear to differ from the type of singularities that the weak cosmic censorship conjecture seeks to prohibit.
Therefore, the evolution of Nariai black holes requires more careful consideration, and our result should be interpreted as reflecting a property peculiar to overcharging processes.   

Finally, we would like to briefly discuss the thought experiments involving black holes that aim to restrict the spectrum of charged particles, such as the weak gravity conjecture.
As we mentioned in the introduction, 
the weak cosmic censorship conjecture is regarded as a guiding principle for such constraints.
Our result can also be regarded as a support of this kind of constraints.
Furthermore, if we can extend the technique of Sorce-Wald formalism to discuss a particle emission process, it could lead to a more rigorous formulation of such constraints. We leave these interesting directions for future work.

\acknowledgments
D.Y. and K.Y. would like to thank Chandrachur Chakraborty, Hideo Furugori, Kazuyuki Furuuchi, Keisuke Izumi, Anshuman Maharana, Kanji Nishii, Toshifumi Noumi, and Sota Sato for their valuable discussions through the JSPS--DST Bilateral Joint Research Project.
D.Y. is also grateful to Ken-ichi Nakao and Hirotaka Yoshino for their helpful comments.
D.Y. is supported by Grant-Aid for Scientific Research from Ministry of Education, Science, Sports and Culture of Japan (No. JP21H05189) and JSPS KAKENHI Grant No. JP20K14469.  D.Y. and K.Y. are supported by Japan–India Cooperative Scientific Programme between JSPS and DST Grant Number JPJSBP120227705.  K.Y is also supported by JST SPRING, Grant Number JPMJSP2108. 

\appendix
\section{Gaussian Null coordinate}
\label{App:GNC}
\subsection{Definition and Basic Property}
In this appendix, we summarize properties of the Gaussian null coordinates \cite{Hollands:2006rj,Hollands:2012sf}. 
For a given null hypersurface $\mathcal{H}$, the Gaussian null coordinates $\{u, v,\theta, \varphi \}$ are defined by  
\begin{align}
    g_{\mu\nu} \mathbf{d}x^\mu \mathbf{d}x^\nu=2(\mathbf{d}v - v^2 \alpha \mathbf{d}u - v \beta_A  \mathbf{d}x^A)\mathbf{d}u+\gamma_{AB} \mathbf{d}x^A \mathbf{d}x^B,\label{GNC}
\end{align}
where $x^A=(\theta,\varphi)$ represent coordinates on the 2-dimensional surfaces $C(u,v)$ of constant $u$ and $v$. Here $\alpha$, $\beta_{A}$, and  $\gamma_{AB}$ are functions of $x^{\mu}$, which are assumed to be finite at $\mathcal{H}$, $v = 0$.

We define the vector fields $\boldsymbol{k}$ by $\boldsymbol{k} \coloneqq \boldsymbol{\partial}_{u}$.
By the definition, $\boldsymbol{k}$ is tangent to the $v$ constant surfaces including ${\cal H}$. The dual vector for $\boldsymbol{k}$ can be expressed as 
 \begin{align}
 k_{\mu} \mathbf{d} x^{\mu} = \mathbf{d} v - 2 v^2 \alpha \mathbf{d} u - v \beta_{A} \mathbf{d} x^{A}  \overset{\mathcal{H}}{=}  \mathbf{d} v.
\end{align}
Hence, $\boldsymbol{k}$ is also a normal vector of $v = 0$,  hypersurface $\mathcal{H}$. The norm of $\boldsymbol{k}$ is evaluated as 
\begin{align}
 k_{\mu} k^{\mu} = - 2 v^2 \alpha \overset{\mathcal{H}}{=} 0.
\end{align}
This represents $\boldsymbol{k}$ is null vector on ${\cal H}$. 
In addition, one can check that $\boldsymbol{k}$ is a tangent vector of affine parametrized null geodesic generators on $\mathcal{H}$, $k^{\mu} \nabla_{\mu} k^{\nu} \overset{\mathcal{H}}{=} 0$, and hence $u$ is the affine parameter of them. We identify the future direction as that of the vector $\boldsymbol{k}$ on ${\cal H}$.
 
We also define the vector field $\boldsymbol{l} = \boldsymbol{\partial}_{v}$.
One can find that $\boldsymbol{l}$ is the tangent vector of affine parametrized geodesics, $l^{\nu}\nabla_{\nu} l^{\mu} = 0$.
Since
\begin{align}
 l_{\mu} l^{\mu} = 0,
\end{align}
and
\begin{align}
 g_{\mu\nu} k^{\mu} l^{\nu} = 1,
\end{align}
$\boldsymbol{l}$ is null and past directed.
The dual vector can be expressed as 
\begin{align}
 l_{\mu} \mathbf{d}x^{\mu} = \mathbf{d} u.
\end{align}
On the horizon, $\boldsymbol{k}$ and $\boldsymbol{l}$ are null vector fields orthogonal to $C(u,0)$. Therefore, the induced metric on $C(u,0)$ can be expressed as
\begin{align}
\gamma_{\mu\nu}
    \overset{\mathcal{H}}{=}  g_{\mu\nu} - k_{\mu} l_{\nu} - l_{\mu} k_{\nu}. 
\end{align}

\subsection{Volume Elements}
In the Gaussian Null coordinates, the volume element $\boldsymbol{\epsilon}$ can be expressed as 
\begin{align}
 \boldsymbol{\epsilon}&=\sqrt{\gamma}\mathbf{d}u\wedge \mathbf{d}v\wedge \mathbf{d}\theta \wedge \mathbf{d}\varphi.
\end{align}
Evaluated on $\mathcal{H}$, it can be expressed as
\begin{align}
    \label{eq:volumeform432}
\boldsymbol{\epsilon}& \overset{\mathcal{H}}{=}
- \sqrt{\gamma}~\boldsymbol{k} \wedge \boldsymbol{l} \wedge \mathbf{d}\theta \wedge \mathbf{d}\varphi\nonumber\\
    &=-\boldsymbol{k} \wedge \boldsymbol{\epsilon}^{(3)}\nonumber\\
    &=-\boldsymbol{k} \wedge\boldsymbol{l}\wedge   \boldsymbol{\epsilon}^{(2)}.
\end{align}
We note that $\boldsymbol{k}$ and $\boldsymbol{l}$ represent $k_\mu \mathbf{d}x^\mu$ and $l_\mu \mathbf{d}x^\mu$ in this expression.
Here, $\boldsymbol{\epsilon}^{(3)}$ and $\boldsymbol{\epsilon}^{(2)}$ are volume elements on $\mathcal{H}$ $(v = 0)$ and 2-surface $C(u,0)$ respectively, which are given by
\begin{align}
\boldsymbol{\epsilon}^{(3)} \coloneqq \sqrt{\gamma} ~ \mathbf{d} u \wedge \mathbf{d} \theta \wedge \mathbf{d} \varphi 
\end{align}
and
\begin{align}
\boldsymbol{\epsilon}^{(2)} \coloneqq \sqrt{\gamma} ~  \mathbf{d} \theta \wedge \mathbf{d} \varphi .
\end{align}
We note that it can be expressed in the component notation as 
\begin{align}
\epsilon_{\mu\nu\rho\sigma} = - 4 k_{[\mu} \epsilon^{(3)}_{\nu\rho\sigma]}
 = - 12 k_{[\mu} l_{\nu} \epsilon^{(2)}_{\rho\sigma]}.
 \label{eq:epsilon432}
\end{align}

Let us derive the formula of Gauss's theorem on a null surface. Let $\mathcal{H}_{1}$ be a portion of $\mathcal {H}$ and $f$ be an arbitrary function. Then, we obtain
\begin{align}
\int_{\mathcal{H}_{1}} \nabla_{\mu} \left( k^{\mu} f \right) \boldsymbol{\epsilon}^{(3)}
&= 
\int_{\mathcal{H}_{1}} \frac{1}{\sqrt{-g}} \partial_{\mu} \left( \sqrt{-g} k^{\mu} f \right) \sqrt{\gamma} \mathbf{d}u \wedge \mathbf{d}\theta \wedge \mathbf{d} \varphi \notag\\
& = \int_{\mathcal{H}_{1}} \frac{\partial}{\partial u} \left( \sqrt{\gamma} f \right) \mathbf{d}u \wedge \mathbf{d}\theta \wedge \mathbf{d} \varphi \notag\\
&= \int_{\partial \mathcal{H}_{1}} \sqrt{\gamma} f \mathbf{d}\theta \wedge \mathbf{d}\varphi  \notag\\
&= \int_{\partial \mathcal{H}_{1}} f \boldsymbol{\epsilon}^{(2)}. \label{eq:Gauss}
\end{align}

\subsection{Pullback to ${\cal H}$ and $C(u,0)$}
For any 3-form $\boldsymbol{\alpha}$, the pullback to $v = 0$ hypersurface $\mathcal{H}$, denoted as $\underline{\boldsymbol{\alpha}}$, can be evaluated as 
\begin{align}
 \underline{\boldsymbol{\alpha}} = \alpha_{u \theta \varphi} \mathbf{d}u \wedge \mathbf{d} \theta \wedge \mathbf{d} \varphi
= \frac{1}{2 !} \alpha_{\alpha\beta\gamma} k^{\alpha} \gamma^{\beta}{}_{\nu} \gamma^{\gamma}{}_{\rho} l_{\mu} \mathbf{d} x^{\mu} \wedge \mathbf{d} x^{\nu} \wedge \mathbf{d} x^{\rho},
\end{align}
Thus, in the component notation, we obtain
\begin{align}
 \underline{\alpha}_{\mu\nu\rho} = 3 \alpha_{\alpha\beta\gamma} k^{\alpha} l_{[\mu}\gamma^{\beta}{}_{\nu} \gamma^{\gamma}{}_{\rho]}.
\end{align}
An important consequence of this expression is,
\begin{align}
 i_{\boldsymbol{k}} \boldsymbol{\alpha}  = 0 ~  \Rightarrow \int_{\mathcal{H}} \boldsymbol{\alpha} = 0. \label{inta=0}
\end{align}
In addition, for $\alpha_{\mu\nu\rho} = V^{\sigma} \epsilon_{\sigma\mu\nu\rho}$, by using Eq.~\eqref{eq:epsilon432}, we obtain $\underline{\alpha}_{\mu\nu\rho} = - V^{\sigma} k_{\sigma} \epsilon^{(3)}_{\mu\nu\rho}$. This indicates
\begin{align}
 \int_{\mathcal{H}} \frac{1}{3!} V^{\sigma} \epsilon_{\sigma\mu\nu\rho} \mathbf{d} x^{\mu} \wedge \mathbf{d} x^{\nu} \wedge \mathbf{d} x^{\rho}
 = - \int_{\mathcal{H}} V^{\sigma} k_{\sigma} \boldsymbol{\epsilon}^{(3)}. \label{eq:intVe} 
\end{align}

Similarly, for any 2-form $\boldsymbol{\beta}$, the pullback to $u =$ constant, $v = 0$ surfaces $C(u,0)$ can be evaluated as 
\begin{align}
 \underline{\boldsymbol{\beta}} = \beta_{\theta \varphi} \mathbf{d} \theta \wedge \mathbf{d} \varphi
= \frac{1}{2!} \beta_{\alpha\beta} \gamma^{\alpha}{}_{\mu} \gamma^{\beta}{}_{\nu} \mathbf{d} x^{\mu} \wedge \mathbf{d} x^{\nu}.
\label{eq:2formpullback}
\end{align}
In the component notation, it can be represented as 
\begin{align}
 \underline{\beta}_{\mu\nu} = \beta_{\alpha\beta} \gamma^{\alpha}{}_{\mu} \gamma^{\beta}{}_{\nu}.
 \label{eq:2formpullbackcomponent}
\end{align}

\subsection{Reissner--Nordstr\"{o}m--de Sitter Spacetime in the Gaussian Null Coordinates}
In this subsection, we explicitly construct the Gaussian null coordinates for static, spherically symmetric class of black hole spacetime
\begin{align}
 \boldsymbol{g} = - f(r) \mathbf{d} t^2 + \frac{\mathbf{d}r^2}{f(r)} + r^2 \left( \mathbf{d} \theta^2 + \sin^2 \theta \mathbf{d} \varphi^2 \right).
\end{align} 
The static Killing vector is 
\begin{align}
 \boldsymbol{\xi} = \boldsymbol{\partial}_{t},
\end{align}
and we assume this spacetime has a Killing horizon at $r = r_{\mathrm{H}}$. 

Let us introduce Eddington--Finkelstein coordinate $U \coloneqq t + r_{*}(r)$, where the tortoise coordinate $r_{*}$ is defined by $r_{*} \coloneqq \int f(r)^{-1} dr$. 
The metric can be expressed as
\begin{align}
 \boldsymbol{g} = - f(r) \mathbf{d} U^2 + 2 \mathbf{d}U \mathbf{d}r + r^2 \left( \mathbf{d} \theta^2 + \sin^2 \theta \mathbf{d} \varphi^2 \right).
\end{align}
The static Killing vector can be expressed as $\boldsymbol{\xi} = \boldsymbol{\partial}_{U}$.
Since $U, \theta, \varphi$ constant curves with a parameter $r$ are past directed null geodesics, we assume the vector field $\boldsymbol{l}$ can be expressed as 
\begin{align}
 \boldsymbol{l} = l^{r}(U) \boldsymbol{\partial}_{r}.
\end{align}
$\boldsymbol{l}$ has desired properties, $l_{\mu} l^{\mu} = 0$ and $l^{\mu} \nabla_{\mu} l^{\nu} = 0$.
Then let us construct the vector field $\boldsymbol{k}$ which has the following properties: (i) $g_{\mu\nu} k^{\mu} l^{\nu} = 1$, (ii) $[\boldsymbol{k},\boldsymbol{l}] = 0$, (iii) $k^{\mu} \nabla_{\mu} k^{\nu} \overset{r = r_{\mathrm{H}}}{=} 0$ and (iv) $k_{\mu} k^{\mu} \overset{r = r_{\mathrm{H}}}{=} 0$.
Starting from the ansatz, 
\begin{align}
 \boldsymbol{k} =  k^{U}(U,r) \boldsymbol{\partial}_{U} + k^{r}(U, r) \boldsymbol{\partial}_{r},
\end{align}
above four conditions determines $l^{r}, k^{U}$ and $k^{r}$ up to a constant scale as follows,
\begin{align}
 \boldsymbol{l} = \mathrm{e}^{\kappa U} \boldsymbol{\partial}_{r}, 
\qquad 
 \boldsymbol{k} = \mathrm{e}^{- \kappa U} \left( \boldsymbol{\partial}_{U} + \kappa (r - r_{\mathrm{H}}) \boldsymbol{\partial}_{r} \right), \label{lkinEFC}
\end{align}
where $\kappa = f'(r_{\mathrm{H}})/2$.
Since the condition (ii) indicates that $\boldsymbol{l}$ and $\boldsymbol{k}$ are coordinate basis, one can introduce new coordinates $u$ and $v$ by
$\boldsymbol{k} = \boldsymbol{\partial}_{u}$ and $\boldsymbol{l} = \boldsymbol{\partial}_{v}$. Comparing the expressions in \eqref{lkinEFC}, the coordinate transformation can be explicitly derived as 
\begin{align}
 U = \frac{1}{ \kappa} \log (\kappa u), \qquad  r = r_{\mathrm{H}} + \kappa u v .
\end{align}
Using the coordinates $\{ u, v, \theta, \varphi \}$, the metric can be expressed as 
\begin{align}
 \boldsymbol{g} = 2 \left( \mathbf{d} v -  v^2 \alpha(u,v) \mathbf{d} u \right)\mathbf{d}u + (r_{\mathrm{H}} + \kappa u v)^2 (\mathbf{d} \theta^2 + \sin^2 \theta \mathbf{d} \varphi^2),
\label{eq:RNdSinGNC}
\end{align}
where the function $\alpha$ is given by
\begin{align}
 \alpha(u,v) \coloneqq \frac{1}{\kappa^2 u^2 v^2} \left( \frac{1}{2}f(r_{\mathrm{H}} + \kappa u v)
 - \kappa^2 u v
\right)
= \frac{1}{4} f''(r_{\mathrm{H}}) + {\cal O}(u v),
\end{align}
and hence it is finite at $v = 0$. 
The expression \eqref{eq:RNdSinGNC} is nothing but the metric in the Gaussian null coordinates.
The null surface $\mathcal{H}$ defined by $v = 0$ with $u \geq 0$ corresponds to the event horizon. The two surfaces $C(u,0)$ defined by $v = 0$ and $u =$ constant are two-dimensional spheres that represent cross-sections of the event horizon. In particular $C(0,0)$ corresponds to the bifurcation surface.

Finally let us check the expression of the static Killing vector $\boldsymbol{\xi}$ in the Gaussian null coordinates. 
It can be obtained as 
\begin{align}
 \boldsymbol{\xi} = \boldsymbol{\partial}_{U}
= \mathrm{e}^{\kappa U} \boldsymbol{k} - \kappa ( r - r_{\mathrm{H}}) \mathrm{e}^{- \kappa U} \boldsymbol{l}
= \kappa ( u \boldsymbol{k} - v \boldsymbol{l}).
\label{eq:GNCkilling}
\end{align}

\subsection{Perturbations}
\label{app:GNCpert}
In the main section, we consider one parameter family of the metric in the Gaussian null coordinates defined by 
\begin{align}
    g_{\mu\nu}(\lambda) \mathbf{d}x^\mu \mathbf{d}x^\nu=2(\mathbf{d}v - v^2 \alpha(\lambda) \mathbf{d}u - v \beta_A(\lambda)  \mathbf{d}x^A)\mathbf{d}u+\gamma_{AB}(\lambda) \mathbf{d}x^A \mathbf{d}x^B.\label{GNClambda}
\end{align}
By the definitions, $\boldsymbol{k} = \boldsymbol{\partial}_{u}$, $\boldsymbol{l} = \boldsymbol{\partial}_{v}$, $l_{\mu} \mathbf{d} x^{\mu} = \mathbf{d} u$ and $\boldsymbol{\xi} = \kappa \left(u\boldsymbol{k}-v\boldsymbol{l}\right)$ are independent of $\lambda$. 
In addition, since $k_{\mu} \mathbf{d}x^{\mu} \overset{\mathcal{H}}{=} \mathbf{d} r$ and $\xi_{\mu} \mathbf{d}x^{\mu} \overset{\mathcal{H}}{=} \kappa u \mathbf{d} r$, $k_{\mu}$ and $\xi_{\mu}$ are also independent of $\lambda$ on $\mathcal{H}$.
In particular, we obtain
\begin{align}
g_{\mu\nu}(\lambda) \xi^{\nu} \overset{\mathcal{H}}{=} \bar{g}_{\mu\nu} \xi^{\nu}. 
\label{xibaronH}
\end{align}
Thus, it does not matter which metric is used to lower the indices of $\xi^{\mu}$, as well as $k^{\mu}$ and $l^{\mu}$, on $\mathcal{H}$.
This property can be rephrased as $\delta g_{\mu\nu}$ is a tensor projected to $C(u,0)$, that is, $\delta g_{\mu\nu}$ satisfies
\begin{align}
 k^{\mu} \delta g_{\mu\nu} \overset{\mathcal{H}}{=} 0, \qquad l^{\mu} \delta g_{\mu\nu} \overset{\mathcal{H}}{=} 0, \label{eq:kdeltag=0}
\end{align}
and
\begin{align}
 \delta g_{\mu\nu}  \overset{\mathcal{H}}{=} \delta \gamma_{\mu\nu}.
\label{deltag=deltagamma}
\end{align}
As a consequence, the Lie derivative can be expressed as 
\begin{align}
 \mathsterling_{\boldsymbol{\xi}} \delta g_{\mu\nu} \overset{\mathcal{H}}{=} \kappa u \mathsterling_{\boldsymbol{k}} \delta g_{\mu\nu} + \kappa \delta g_{\rho\nu} l_\mu k^{\rho} + \kappa \delta g_{\mu\rho} l_\nu k^{\rho}
\overset{\mathcal{H}}{=} \kappa u \mathsterling_{\boldsymbol{k}} \delta g_{\mu\nu}. \label{eq:Liexi=Liek}
\end{align}

Let us define the extrinsic curvature $K_{\mu\nu}$ associated with $\boldsymbol{k}$ on the $u =$ constant, $v = 0$ 2-surface $C(u,0)$ by
\begin{align}
 K_{\mu\nu} = \gamma_{\mu}{}^{\rho} \gamma_{\nu}{}^{\sigma} \nabla_{\rho} k_{\sigma} = \frac{1}{2} \mathsterling_{\boldsymbol{k}} \gamma_{\mu\nu}.
\end{align}
Then, we define the expansion $\vartheta$ and shear $\sigma_{\mu\nu}$ by
\begin{align}
 \vartheta &\coloneqq K^{\mu}{}_{\mu}, \\
 \sigma_{\mu\nu} &\coloneqq K_{\mu\nu} - \frac{1}{2} \vartheta \gamma_{\mu\nu}. 
\end{align}
Since $\boldsymbol{k}$ is the Horizon generator for the background static spacetime, one can see that 
\begin{align}
 \bar{\vartheta} &= \bar{\nabla}_\mu k^\mu= 0 
  \label{eq:barsigma}\\
 \bar{\sigma}_{\mu\nu} &= 0 ~.
\end{align}
The first order perturbations of $K_{\mu\nu}$ can be expressed as 
\begin{align}
 \delta K_{\mu\nu} = \frac{1}{2} \mathsterling_{\boldsymbol{k}} \delta \gamma_{\mu\nu} = \frac{1}{2} \mathsterling_{\boldsymbol{k}} \delta g_{\mu\nu} = \frac{1}{2\kappa u} \mathsterling_{\boldsymbol{\xi}} \delta g_{\mu\nu}, 
\end{align}
and hence the first order perturbations of expansion $\vartheta$ and shear $\sigma_{\mu\nu}$ can be expressed as
\begin{align}
 \delta \vartheta = \bar{\gamma}^{\mu\nu} \delta K_{\mu\nu}
 =  \frac{1}{2 \kappa u} \bar{g}^{\mu\nu} \mathsterling_{\boldsymbol{\xi}} \delta g_{\mu\nu} = \frac{1}{2 \kappa u} \mathsterling_{\boldsymbol{\xi}} (\bar{g}^{\mu\nu} \delta g_{\mu\nu}) = \frac{1}{2} \mathsterling_{\boldsymbol{k}} \left(\bar{g}^{\mu\nu} \delta g_{\mu\nu} \right). 
\end{align}
Here we use $\bar{K}_{\mu\nu} = 0$ in the first equality and $\mathsterling_{\boldsymbol{\xi}} \bar{g}^{\mu\nu} = 0$ in the third equality.

Since $\boldsymbol{k}$ is null geodesic on $\mathcal{H}$, it satisfies the Raychaudhuri equations,
\begin{align}
    \frac{d}{du}\vartheta &\overset{\mathcal{H}}{=} -\frac{1}{2}\vartheta^2-\sigma_{\mu\nu} \sigma^{\mu\nu} - R_{\mu\nu} k^\mu k^\nu.
\end{align}
If we take the $\lambda$ derivative of the Raychaudhuri equation and set $\lambda=0$, it becomes
\begin{align}
    \frac{d}{du}\delta\vartheta \overset{\mathcal{H}}= - \bar{\vartheta} \delta\vartheta  - 2 \bar{\sigma}^{\mu}{}_{\nu} \delta \sigma^{\nu}{}_{\mu}  - \delta R_{\mu\nu} k^\mu k^\nu \overset{\mathcal{H}} = - \delta R_{\mu\nu} k^\mu k^\nu.
\end{align}
When the null energy condition for the first order perturbations are saturated on $\mathcal{H}$, as assumed in Eq.~\eqref{eq:optimalNEC} in the main section, $\delta R_{\mu\nu} k^{\mu} k^{\nu}|_{\mathcal{H}}$ vanishes. Thus $\delta \vartheta$ is constant on $\mathcal{H}$. Combining it with the assumption where the perturbation vanishes at the bifurcation surface $B$ as $\left. \delta \vartheta \right|_{B}=0$, we find that  $\delta \vartheta = 0 $.
This implies 
\begin{align}
k^{\rho} \bar{\nabla}_{\rho} \left(\bar{g}^{\mu\nu} \delta g_{\mu\nu} \right)
= \mathsterling_{\boldsymbol{k}} \left(\bar{g}^{\mu\nu}\delta g_{\mu\nu}\right) \overset{\mathcal{H}}{=} 0, \qquad \mathsterling_{\boldsymbol{\xi}} \left(\bar{g}^{\mu\nu} \delta g_{\mu\nu} \right) \overset{\mathcal{H}}{=} 0.
 \label{eq:Liektrdeltag}
\end{align}
We use these property in the evaluation of Eq.~\eqref{inthesquarebracket}.
Note that $\delta \vartheta = 0$ indicates
\begin{align}
    \delta \sigma_{\mu\nu} = \frac{1}{2}\mathsterling_{\boldsymbol{k}} \delta g_{\mu\nu}. \label{deltasigma=}
\end{align}
Other properties used in the derivation of Eq.~\eqref{inthesquarebracket} are $k^\mu \mathsterling_{\boldsymbol{\xi}}\delta g_{\mu\nu}=0$ and $\bar{g}^{\mu\nu} k^\rho \bar{\nabla}_\mu \mathsterling_{\boldsymbol{\xi}} \delta g_{\nu\rho} = 0$. The former property can be shown as 
\begin{align}
 k^\mu \mathsterling_{\boldsymbol{\xi}}\delta g_{\mu\nu}
= \kappa u k^{\mu} \mathsterling_{\boldsymbol{k}} \delta g_{\mu\nu}
= \kappa u \mathsterling_{\boldsymbol{k}} \left(k^{\mu} \delta g_{\mu\nu} \right)  \overset{\mathcal{H}}{=} 0. \label{eq:kLiedeltag=0}
\end{align}
Here we use Eq.~\eqref{eq:Liexi=Liek} in the first equality and $\mathsterling_{\boldsymbol{k}} \boldsymbol{k} = 0$ in the second equality.
The latter property can be derived as 
\begin{align}
\bar{g}^{\mu\nu} k^\rho \bar{\nabla}_\mu \mathsterling_{\boldsymbol{\xi}} \delta g_{\nu\rho} \overset{\mathcal{H}}{=} - \bar{g}^{\mu\nu} \bar{\nabla}_\mu  k^\rho  \mathsterling_{\boldsymbol{\xi}} \delta g_{\nu\rho}
\overset{\mathcal{H}}{=} - \bar{\gamma}^{\mu\nu} \bar{\nabla}_\mu  k^\rho  \mathsterling_{\boldsymbol{\xi}} \delta \gamma_{\nu\rho} = -\bar{K}^{\nu\rho} \mathsterling_{\boldsymbol{\xi}} \delta \gamma_{\nu\rho} \overset{\mathcal{H}}{=} 0, \label{eq:kdelLiedeltag}
\end{align}
where the first and the second equalities follows from Eqs.~\eqref{eq:kdeltag=0} and \eqref{eq:kLiedeltag=0}, respectively.

\section{Relation to the Second Law}
\label{App:2nd}
In this section, we clarify the relation between our analysis and the requirement of the second law of black hole thermodynamics along the studies in Refs.~\cite{Lin:2022ndf, Lin:2024deg, Wu:2024ucf}.
Let us consider one parameter family of Reissner--Nordstr\"{o}m--de Sitter black holes
 with the mass parameter $M(\alpha)$ and the charge $Q(\alpha)$ as
 \begin{align}
    M(\alpha)=\bar{M}+\delta M\alpha+\frac{1}{2}\delta^2M\alpha^2+\cdots, \quad Q(\alpha)=\bar{Q}+\delta Q\alpha+\frac{1}{2}\delta^2Q\alpha^2+\cdots.
\end{align}
The function $f(r)$ in the metric of the Reissner--Nordstr\"{o}m--de Sitter black hole is given by
\begin{align}
 f(r; M(\alpha), Q(\alpha), H) = 1 - \frac{2 G M(\alpha)}{r} + \frac{G k Q(\alpha)^2}{r^2} - H^2 r^2 .
\end{align}
The position of the horizon for a black hole with $\alpha$ can be determined by
\begin{align}
 f(r_{\mathrm{H}}(\alpha); M(\alpha), Q(\alpha), H) = 0,
\end{align}
where $r_{\mathrm{H}}(\alpha)$ is also expanded as
\begin{align}
    r_{\mathrm{H}}(\alpha) = \bar{r}_{\mathrm{H}}+\delta r_{\mathrm{H}}\alpha+\frac{1}{2}\delta^2r_{\mathrm{H}}\alpha^2+\cdots
\end{align}
By expanding above equation with respect to $\alpha$ and evaluating order by order, we obtain
\begin{align}
\alpha^0:~ &f(\bar{r}_{\mathrm{H}}; M, Q, H)=0 \notag\\
\alpha^1:~ &f_{r} \delta r_{\mathrm{H}} + f_{M} \delta M + f_{Q} \delta Q  =0\notag\\
\alpha^2:~ &\frac{1}{2} \Bigl(
f_{r} \delta^2 r_{\mathrm{H}} + f_{M} \delta^2 M + f_{Q} \delta^2 Q \notag\\
&~~+ f_{rr} \delta r_{\mathrm{H}}^2
+ f_{MM} \delta M^2
+ f_{QQ} \delta Q^2
+ 2 f_{M r} \delta r_{\mathrm{H}} \delta M 
+ 2 f_{Q r} \delta r_{\mathrm{H}} \delta Q
\Bigr)=0 ,
\end{align}
where a subscription of $f$ represents the partial derivative of $f$ with respect to the subscript as $f_M=\frac{\partial f}{\partial M}$ and so on.
Solving these equations, $\delta r_{\mathrm{H}}$ and $\delta^2 r_{\mathrm{H}}$ are expressed in terms of the deviations of the mass and the charge as, 
\begin{align}
 \delta r_{\mathrm{H}} &= - \frac{f_{M}}{f_{r}} \left( \delta M + \frac{f_{Q}}{f_{M}} \delta Q \right), \label{eq:deltarH}\\
 \delta^2 r_{\mathrm{H}} &= - \frac{f_{M}}{f_{r}} \left( \delta^2 M + \frac{f_{Q}}{f_{M}}\delta^2 Q  \right) + (...) \delta M^2 + (...) \delta M \delta Q + (...) \delta Q^2~.\label{eq:delta2rH}
\end{align}
On other hand, the area of the Reissner--Nordstr\"{o}m--de Sitter black hole is expanded as
\begin{align}
 A^{\mathrm{RNdS}}(\alpha) \coloneqq 4 \pi r_{\mathrm{H}}(\alpha)^2
= \bar{A}^{\mathrm{RNdS}} + \delta A^{\mathrm{RNdS}} \alpha + \frac{1}{2} \delta^2 A^{\mathrm{RNdS}} \alpha^2 + {\cal O}(\alpha^3),
\end{align}
where the deviations of the area are obtained as
\begin{align}
 \bar{A}^{\mathrm{RNdS}} &= 4 \pi \bar{r}_{\mathrm{H}}^2 \\
 \delta A^{\mathrm{RNdS}} &= 8 \pi \bar{r}_{\mathrm{H}} \delta r_{\mathrm{H}} \\
 \delta^2 A^{\mathrm{RNdS}} &= 8 \pi (\bar{r}_{\mathrm{H}} \delta^2 r_{\mathrm{H}}  +  \delta r_{\mathrm{H}}^2)~. \label{eq:delta2A}
\end{align}
By substituting Eq.~\eqref{eq:delta2A} for Eq.~\eqref{eq:deltarH} and Eq.~\eqref{eq:delta2rH}, 
\begin{align}
 \delta^2 A^{\mathrm{RNdS}} &= - \frac{8 \pi f_{M}}{f_{r}}\bar{r}_{\mathrm{H}} \left(\delta^2 M + \frac{f_{Q}}{f_{M}} \delta^2 Q \right) 
+ \delta^2 A^{\mathrm{RNdS}}|_{\delta^2 M = \delta^2 Q = 0}.
\end{align}
where all the terms of $\delta M^2$, $\delta Q^2$, and $\delta M \delta Q$ are included in the second term. We can deform this by using a surface gravity $\kappa = \frac{1}{2} f_{r}$ and the potential $\Phi_{\mathrm{H}}=\frac{k \bar{Q}}{\bar{r}_{\mathrm{H}}}$ as
\begin{align}
 \frac{\kappa}{8 \pi G} \delta^2 A^{\mathrm{RNdS}}
&= 
- \frac{f_{M}}{2 G} \bar{r}_{\mathrm{H}} \left( \delta^2 M + \frac{f_{Q}}{f_{M}} \delta^2 Q \right)
 + \frac{\kappa}{8 \pi G} \delta^2 A^{\mathrm{RNdS}}|_{\delta^2 M = \delta^2 Q = 0} \notag\\
&=  \delta^2 M - \Phi_{\mathrm{H}} \delta^2 Q  + \frac{\kappa}{8 \pi G} \delta^2 A^{\mathrm{RNdS}}|_{\delta^2 M = \delta^2 Q = 0}.
\end{align}
The second law approach requires $\delta^2 A^{\mathrm{RNdS}} \geq 0$, which reduces to
 \begin{align}
  \delta^2 M - \Phi_{\mathrm{H}} \delta^2 Q \geq - \frac{\kappa}{8 \pi G} \delta^2 A^{\mathrm{RNdS}}|_{\delta^2 M = \delta^2 Q = 0},
 \end{align}
that exactly coincides with the inequality 
\eqref{eq:2nd_MQA} obtained by Sorce--Wald formalism from the null energy condition.

\bibliography{ref} 
\bibliographystyle{JHEP.bst}
\end{document}